\documentclass{article}

\usepackage{PRIMEarxiv}
\usepackage[utf8]{inputenc} 

\usepackage{amsmath,amsfonts,amssymb}
\usepackage{mathtools}

\usepackage{subcaption}

\usepackage{booktabs}
\usepackage{siunitx}
\usepackage{cite}
\usepackage{lettrine}
\usepackage[hidelinks]{hyperref}
\usepackage{comment}
\usepackage{float}
\newfloat{algorithm}{t}{lop}
\usepackage{array}
\usepackage{booktabs}
\usepackage{url}
\usepackage{xcolor}
\usepackage{graphicx}
\usepackage{multirow}
\usepackage{makecell}
\usepackage{caption}
\usepackage[euler]{textgreek}
\usepackage{relsize}
\usepackage{amssymb}
\usepackage{bm}

\usepackage{dblfloatfix}
\usepackage[subnum]{cases}
\usepackage{algorithm}
\usepackage{algpseudocode}

\usepackage{physics}

\renewcommand{\mathbf}{\boldsymbol}

\usepackage{float}
\usepackage{capt-of}
\usepackage{tabularx}
\newcount\Comments  
\usepackage{color}
\definecolor{darkgreen}{rgb}{0,0.5,0}
\definecolor{purple}{rgb}{1,0,1}
\definecolor{amber}{rgb}{1,0.49,0}
\definecolor{uglycolor}{rgb}{0.5,0.5,0}
\newcommand{\kibitz}[2]{\ifnum\Comments=0\textcolor{#1}{#2}\fi}

\makeatletter
\newcommand*{\rom}[1]{\expandafter\@slowromancap\romannumeral #1@}
\makeatother

\title{TRACER: An RDC-Triggered Trajectory Recovery Framework for Cyberattack-Resilient Adaptive Cruise Control
\thanks{\textit{\underline{Citation}}: 
\textbf{Yang et al. TRACER: An RDC-Triggered Trajectory Recovery Framework for Cyberattack-Resilient Adaptive Cruise Control.}} 
}

\author{
  Yicheng Yang \\
  School of Artificial Intelligence \\
  Hebei University of Technology\\
  \AND
  Tianyu Liu \\
  School of Artificial Intelligence \\
  Hebei University of Technology\\
  \AND
  Omar Tahri \\
  Department of Computer Science\\
  Saint Louis University \\
    \AND
    Haoli Zhao \\
  School of Artificial Intelligence \\
  Hebei University of Technology\\
    \AND
    Yongfeng Dong \\
  School of Artificial Intelligence \\
  Hebei University of Technology\\
  \And
  Tianyi Li\thanks{Corresponding author.} \\
  Department of Civil Engineering \\
  Saint Louis University \\
  \texttt{tianyili.ai@gmail.com} 
  }

\begin{document}
\maketitle


\begin{abstract}

The increasing connectivity of adaptive cruise control (ACC) systems exposes vehicle-state information to cyberattacks, potentially causing irrational car-following behavior and unsafe longitudinal dynamics. 
Existing trajectory recovery methods generally lack an interpretable criterion for recovery activation and remain vulnerable to attack-induced distribution shifts. To address these challenges, this paper proposes TRACER, a rational driving constraint (RDC)-triggered trajectory recovery framework for cyberattack-resilient ACC systems. 
TRACER employs a rationality-aware triggering (RAT) mechanism that activates recovery when the observed driving state violates the calibrated RDC bounds. It also introduces cross-branch robust imitation learning (CRIL), which combines a clean branch for learning nominal driving behavior with a noise-enhanced branch for improving robustness to corrupted observations. Cross-branch fine-tuning further aligns the robust recovery
policy with the nominal driving-state distribution. Experiments across six cyberattack scenarios show that TRACER achieves reliable recovery triggering and outperforms state-of-the-art recovery baselines in terms of accuracy, stability, and safety. State-space analysis further confirms that TRACER mitigates attack-induced distribution shifts and restores coupled spacing–velocity dynamics, highlighting the effectiveness of robust policy learning for cyberattack-resilient ACC.

\end{abstract}

\keywords{Car-following \and Adaptive cruise control (ACC) \and adversarial imitation learning \and cyberattack resilience \and artificial intelligence (AI) \and traffic flow \and trajectory recovery  }

\section{Introduction}

Automated vehicles (AVs) have the potential to improve traffic stability, energy efficiency, and mobility \cite{wang2022optimal,sun2022energy}. These benefits increasingly rely on connected cyber-physical functions, including multimodal perception, vehicle-to-everything communication, and advanced driver-assistance systems. However, greater connectivity also expands the attack surface of automated driving systems \cite{yousseef2024autonomous}. Adaptive cruise control (ACC) is particularly vulnerable since its longitudinal control continuously depends on vehicle-state information, including spacing, speed, and relative speed.

Cyberattacks that corrupt such information can alter the acceleration response of an ACC-equipped vehicle and disrupt normal car-following behavior, potentially causing speed oscillations, unsafe inter-vehicle spacing, and traffic instability \cite{kaasen2022towards}.
Therefore, improving cyberattack resilience requires more than detecting or suppressing corrupted signals. Once the vehicle deviates from normal operation, an effective response should activate recovery at an appropriate time and guide the compromised trajectory back toward rational car-following behavior \cite{el2025driving}.


Existing trajectory recovery approaches have primarily focused on reconstruction of compromised trajectories. Sequence-based methods \cite{park2018sequence} commonly formulate recovery as trajectory or velocity reconstruction from historical observations. However, accurate velocity reconstruction alone does not guarantee spacing recovery because spacing evolves through the accumulated relative motion between consecutive vehicles. Generative methods can learn complex trajectory distributions \cite{fu2024trajectory}, but statistical similarity does not necessarily ensure rational car-following behavior. Imitation learning methods derive driving policies directly from expert demonstrations \cite{bhattacharyya2022modeling,eraqi2022dynamic}, yet reproducing nominal behavior alone may be insufficient when recovery must operate on attack-corrupted observations. These limitations indicate that effective trajectory recovery should restore the coupled longitudinal driving state rather than merely reconstruct individual trajectory variables.

In addition to improving recovery of compromised trajectories, a remaining challenge is determining when recovery should be activated. Existing approaches generally lack a physically interpretable criterion for determining whether the observed driving state has departed sufficiently from normal car-following behavior to require recovery \cite{li2025racer,boddupalli2022resilient}. Prediction or reconstruction errors alone are inadequate for this purpose because a trajectory may exhibit a small velocity error while still violating reasonable spacing relationships or longitudinal driving conditions. Rational driving constraints (RDCs) characterize physically meaningful sensitivity relationships underlying car-following dynamics \cite{wang2024string,li2025racer} and have been used to evaluate driving rationality under cyberattacks \cite{wang2023optimal,wang2024analytical}. They therefore provide a physically interpretable basis for triggering trajectory recovery when the observed driving state departs from the admissible car-following region.

A further challenge arises from the distribution shift between nominal expert trajectories and attack-corrupted observations \cite{geng2023adaptive}. As illustrated in Fig. \ref{fig:traditional}, different cyberattacks can drive the vehicle away from the nominal spacing–speed relationship toward distinct abnormal state distributions. Such shifts may limit the generalizability of recovery policies trained
primarily on nominal demonstrations and degrade their performance across attack scenarios \cite{eraqi2022dynamic}. Consequently, an effective recovery framework should not only trigger recovery upon departures from rational car-following behavior but also robustly restore the coupled longitudinal dynamics under attack-induced distribution shifts.

To address these challenges, this paper proposes TRACER, an RDC-triggered trajectory recovery framework for cyberattack-resilient ACC systems. TRACER integrates rationality-aware triggering (RAT) with cross-branch robust imitation learning (CRIL) to activate recovery upon RDC violations and restore attack-compromised trajectories under distribution shifts.
The main contributions of this paper are summarized as follows:

\begin{figure}[t]
    \centering
    \begin{subfigure}{0.31\columnwidth}
        \centering
        \includegraphics[width=\linewidth]{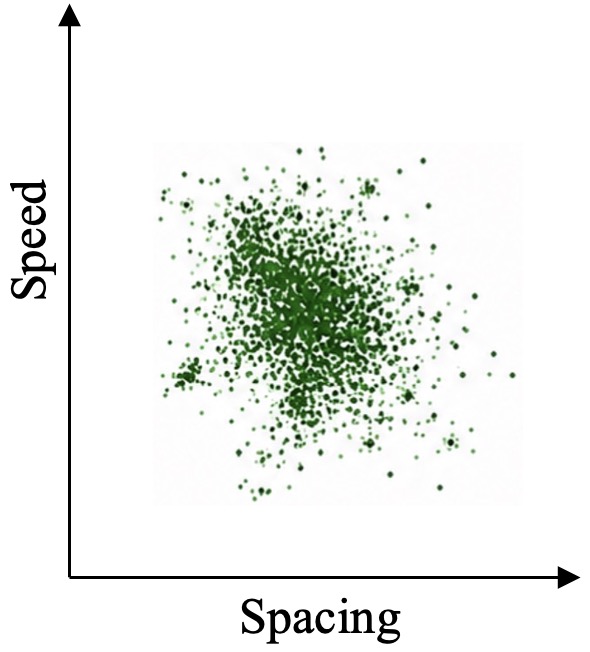}
        \caption{}
        \label{fig:a}
    \end{subfigure}
    \hfill
    \begin{subfigure}{0.30\columnwidth}
        \centering
        \includegraphics[width=\linewidth]{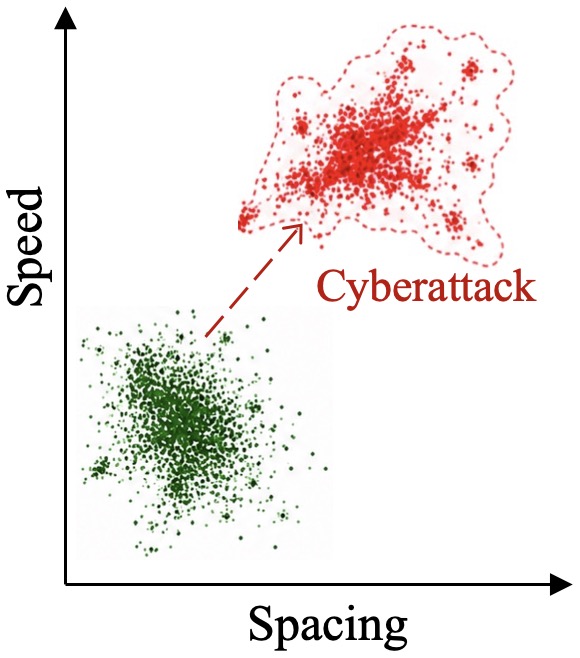}
        \caption{}
        \label{fig:b}
    \end{subfigure}
    \hfill
    \begin{subfigure}{0.33\columnwidth}
        \centering
        \includegraphics[width=\linewidth]{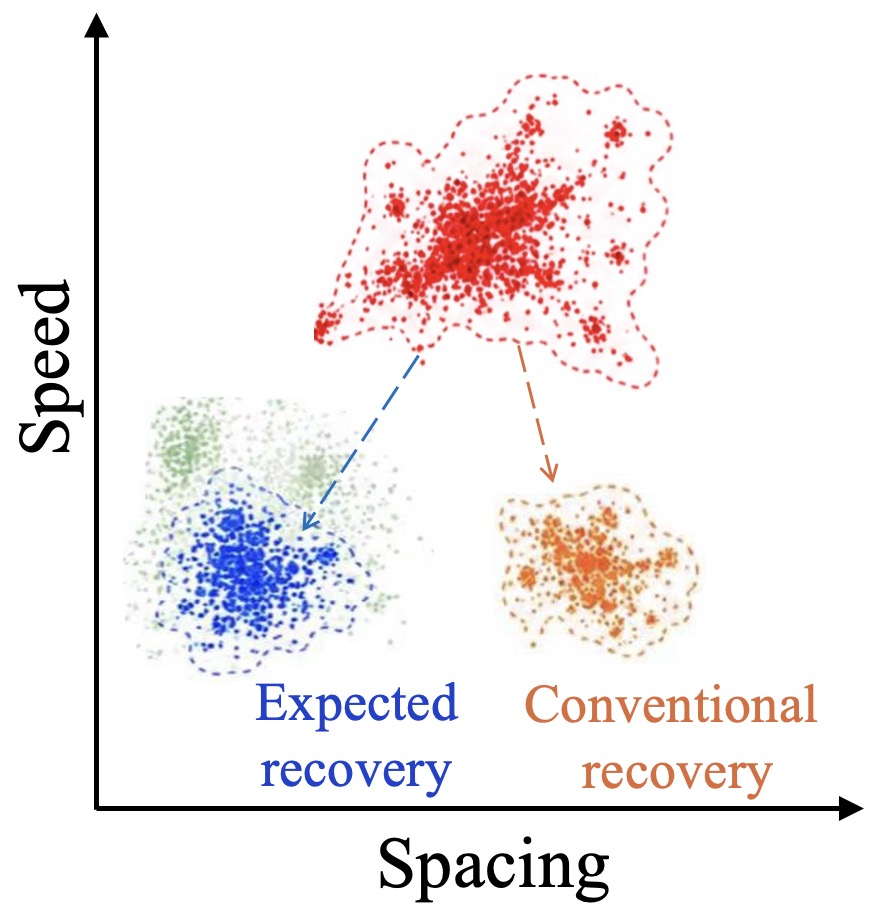}
        \caption{}
        \label{fig:c}
    \end{subfigure}

    \caption{Attack-induced distribution shifts and trajectory recovery in the spacing–speed state space: (a) nominal driving distribution, (b) distinct abnormal distributions induced by cyberattacks, and (c) comparison of recovery outcomes toward the nominal driving-state domain.}
    \label{fig:traditional}
\end{figure}

\begin{itemize}
\item 
We propose TRACER, a unified framework that integrates rationality-aware recovery triggering and robust trajectory recovery for cyberattack-resilient ACC systems.

\item 

We develop RAT, which calibrates the admissible RDC boundaries from attack-free trajectories and activates recovery when the observed driving state violates rational car-following conditions. This mechanism provides a physically interpretable triggering criterion grounded in longitudinal driving dynamics.

\item 


We develop CRIL to mitigate attack-induced distribution shifts. It reconstructs nominal states from corrupted histories and combines clean and noise-enhanced imitation branches to preserve nominal driving behavior and improve robustness, while cross-branch fine-tuning aligns the policy with the nominal driving-state distribution.

\end{itemize}

The remainder of this paper is organized as follows. Section \ref{related works} reviews related work on cyberattack resilience. Section \ref{methodology} introduces the proposed TRACER framework and its mathematical formulation. Section \ref{experiment} presents the experimental setup and comprehensive evaluation under diverse cyberattack scenarios. Section \ref{future research} discusses the limitations of this study. Finally, Section \ref{conclusion} concludes this paper.

\section{Related Work}\label{related works}

Existing cyberattack-response studies can be broadly organized by their primary objectives into attack detection and state estimation, impact mitigation and resilient control, and explicit trajectory recovery. Attack detection and state estimation have been extensively investigated for connected and automated vehicles \cite{ju2022survey,yousseef2024autonomous}. Signal- and model-based approaches employ dynamic watermarking, probing signals, residual analysis, or consistency checks to identify abnormal behavior \cite{porter2019simulation,ibrahum2024deep}. Observer-based methods, including Kalman filtering and sliding-mode observers, further estimate hidden or corrupted states from system dynamics \cite{mousavinejad2018cyber,shin2017illusion}. These approaches are primarily designed to detect information corruption or infer compromised states, but their effectiveness may depend on system models and can degrade under stealthy or complex disturbances.

A separate line of research focuses on impact mitigation and resilient control, aiming to maintain acceptable system operation despite corrupted information. Adversarial training and robust learning improve controller tolerance to perturbations \cite{alotaibi2023adversarial,lin2024safety}, while cooperative perception and trust-weighting mechanisms reduce the influence of unreliable information sources \cite{yu2024online}. Reinforcement-learning-based and other resilience-oriented methods have also been investigated to maintain safety and control performance under cyberattacks \cite{raio2023reinforcement,he2024trustworthy}. Although some of these techniques can support post-attack recovery, their primary objective is to limit attack impact and sustain system operation rather than explicitly restoring compromised driving states.

Trajectory recovery addresses this complementary objective by explicitly reconstructing attack-compromised vehicle states. Existing approaches can be broadly categorized into sequence-based, generative, and imitation learning methods \cite{wang2017capturing,bhattacharyya2022modeling}. Sequence models, such as LSTM \cite{hochreiter1997long} and Transformer \cite{vaswani2017attention}, learn temporal dependencies from historical observations to reconstruct or predict vehicle states. However, optimizing state-level prediction accuracy does not necessarily restore the coupled spacing–velocity dynamics underlying car-following behavior. Generative approaches, particularly diffusion-based models, learn complex driving-state distributions through iterative denoising \cite{yang2024diffusion,jiang2023motiondiffuser}, yet distributional fidelity alone does not ensure rational longitudinal behavior.

Imitation learning directly learns driving policies from expert demonstrations and provides stronger behavioral consistency than state reconstruction \cite{lu2023imitation}. 
Representative paradigms include behavioral cloning \cite{codevilla2019exploring}, inverse reinforcement learning \cite{ng2000algorithms}, and adversarial imitation learning \cite{yun2022parallelized,kuefler2017imitating}. 
Recent adversarial approaches, such as GAIL \cite{ho2016generative}, H-GAIL \cite{couto2024hierarchical}, and DRAIL \cite{lai2024diffusion}, further enhance policy learning through adversarial training, hierarchical structures, and diffusion-based reward modeling. Nevertheless, policies learned primarily from nominal expert demonstrations may encounter distribution shifts when applied to attack-corrupted observations, limiting their robustness across attack conditions.

Existing studies have progressed from attack detection and impact mitigation toward explicit trajectory recovery. However, two challenges remain insufficiently addressed: physically
interpretable recovery triggering and robustness to attack-induced distribution shifts. These limitations motivate the development of TRACER.

\begin{table*}[!t]
\centering
\caption{Summary of Key Notations}
\label{tab:notation}
\resizebox{\textwidth}{!}{
\begin{tabular}{c c c l}
\hline
Symbol & Type & Range & Description \\

\hline

$a_i$, $\hat{a}_i$
& Dynamic variable
& $\mathbb{R}$ 
& 
Nominal and attack-affected accelerations \\

$v$, $\tilde{v}$, $v^{r}$
& State variable
& $(0,+\infty)$
& Vehicle speed under nominal, attacked, and recovered conditions\\





$s$, $\tilde{s}$, $s^{r}$
& State variable
& $(0,+\infty)$
& Vehicle spacing under nominal, attacked, and recovered conditions\\



$\Delta v$, $\Delta\tilde{v}$, $\Delta{v}^r$
& State variable
& $\mathbb{R}$
& Relative speed under nominal, attacked, and recovered conditions \\




$\boldsymbol{\theta}$
& Parameter
& $\mathbb{R}^{6}$
& Calibrated IDM parameters
\\

$\beta_1$
& RDC variable
& $(0,+\infty)$
& Acceleration sensitivity to spacing
\\

$\beta_2$
& RDC variable
& $(0,+\infty)$
& Acceleration sensitivity to relative speed
\\

$\beta_3$
& RDC variable
& $(-\infty,0)$
& Acceleration sensitivity to speed
\\


$\beta_k^{-}$, $\beta_k^{+}$
& Boundary parameter
& $\mathbb{R}$
& Lower and upper boundaries of the $k$-th RDC interval

\\


$\mathcal{C}$
& Constraint set
& $\mathcal{C}_1\times\mathcal{C}_2\times\mathcal{C}_3$
& Admissible RDC region \\

$\Gamma(t)$
& Binary signal
& $\{0,1\}$
& RAT triggering indicator \\

\hline

$O_t^{obs}$
& State vector
& $\mathbb{R}^{3}$
& Observed attacked state \\

$O_t^{E}$
& State vector
& $\mathbb{R}^{3}$
& Attack-free expert state \\

$O_t^{r}$
& State vector 
& $\mathbb{R}^{3}$ 
& Recovered state \\

$H_t^p$, $H_t^{obs}$
& State sequence
& $\mathbb{R}^{W \times 3}$
& Perturbed and corrupted historical windows\\

$W$
& Parameter
& $\mathbb{Z}^{+}$
& Historical window length \\

$\bar{O}_t$
& State vector
& $\mathbb{R}^{3}$
& Reconstructed clean state generated by $G_\psi$ \\


$\hat{O}_t^{E}$, $\hat{O}_t$
& State vector
& $\mathbb{R}^{3}$
& Noise-enhanced expert and reconstructed states \\




$u_t$, $u_t^E$
& Action
& $(0,+\infty)$
& Recovery and expert actions represented by next-step velocities
\\

$G_{\psi}$
& Network
& --
& State reconstruction generator
\\


$\phi_1,\phi_2,\phi_2^{*}$
& Parameter
& --
& Policy network parameters \\

$\pi_{\phi_1}$, $\pi_{\phi_2}$, $\pi{*}$
& Policy
& --
& Clean-branch, noisy-branch, and final recovery policies
\\

$D_1$, $D_2$
& Network
& $(0,1)$
& Clean- and noisy-branch discriminators \\

$\rho_{\pi_{\phi_1}}$, $\rho_{\pi_{\phi_2}}$
& Distribution
& --
& Occupancy distributions of clean and noisy policies \\

\hline
\end{tabular}
}
\end{table*}

\section{Methodology}\label{methodology}

To recover attack-compromised ACC trajectories, we propose TRACER, an RDC-triggered trajectory recovery framework comprising RAT and CRIL, as detailed in the following subsections. RAT evaluates the rationality of the observed driving state and activates recovery upon RDC violations, while CRIL reconstructs compromised trajectories by learning robust recovery policies from attack-free expert demonstrations.

\begin{figure*}[htbp]
\centering
\includegraphics[width=\textwidth]{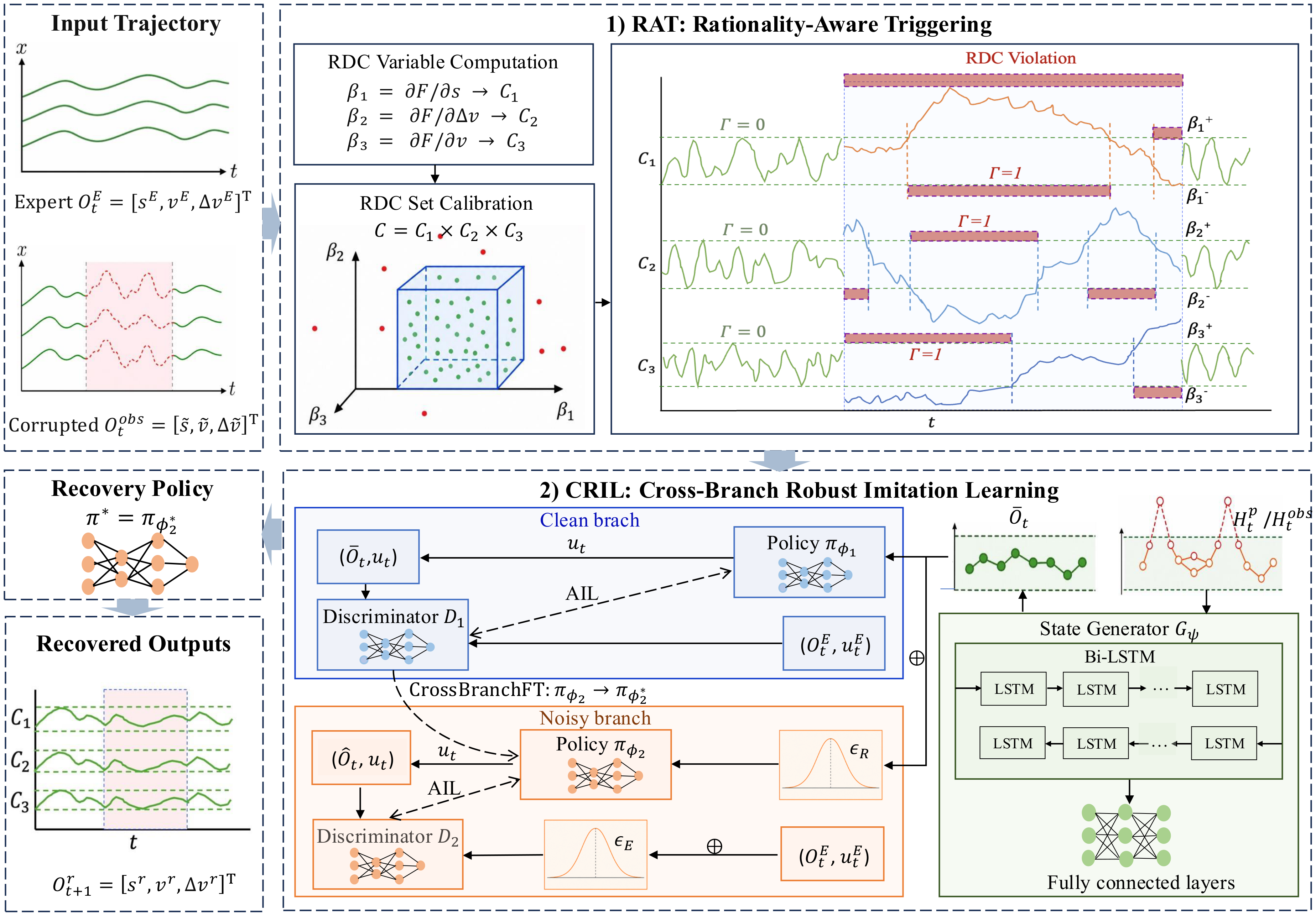}
\caption{
Architecture of the proposed TRACER framework. RAT triggers
recovery when an RDC violation occurs, and CRIL reconstructs
compromised trajectories to follow the nominal driving-state distribution.  
}
\label{fig:framework}
\end{figure*}

For clarity, the key notations used throughout the proposed framework are summarized in Table~\ref{tab:notation}. The overall architecture of TRACER is illustrated in Fig. \ref{fig:framework}, and the main procedures for the training and recovery phases are presented in Algorithms \ref{alg:TRACER_training} and \ref{alg:TRACER_testing}, respectively.

\subsection{Rationality-Aware Triggering}

Microscopic car-following models provide a fundamental framework for describing vehicle interactions \cite{li2024can}. 
Let vehicle \(i\) denote the following vehicle and vehicle \(i-1\) denote its leader. 
At time step \(t\), the spacing, speed, and relative speed are defined as:
\begin{equation}\label{eq:spacing}
s_i(t)=x_{i-1}(t)-x_i(t),
\end{equation}
\begin{equation}\label{eq:speed}
v_i(t)=\dot{x}_i(t),
\end{equation}
\begin{equation}\label{eq:relative speed}
\Delta v_i(t)=v_{i-1}(t)-v_i(t).
\end{equation}
The longitudinal acceleration of vehicle \(i\) is modeled as:
\begin{equation}\label{eq:acceleration}
a_i(t)
=
F\!\left(
\boldsymbol{\theta}_i,
s_i(t),
v_i(t),
\Delta v_i(t)
\right).
\end{equation}
For simplicity, the subscript \(i\) and time argument \(t\) are omitted in the following derivations when no ambiguity arises.

To characterize car-following behavior, this paper adopts the intelligent driver model (IDM) due to its ability to capture acceleration, deceleration, and desired-spacing regulation \cite{talebpour2016influence}. The IDM acceleration function is given by:
\begin{equation}
F(\boldsymbol{\theta},s,v,\Delta v)
=
\zeta
\left[
1-
\left(\frac{v}{v_0}\right)^\delta
-
\left(\frac{s^*(v,\Delta v)}{s}\right)^2
\right],
\end{equation}
where the IDM parameter vector 
\(\boldsymbol{\theta}=[\zeta,b,\delta,s_0,T,v_0]^\top\) 
contains the maximum acceleration \(\zeta\), comfortable deceleration \(b\), acceleration exponent \(\delta\), minimum spacing \(s_0\), desired time gap \(T\), and desired velocity \(v_0\). 
The desired spacing is defined as:
\begin{equation}\label{eq:desired_spacing}
s^*(v,\Delta v)
=
s_0+vT-\frac{v\Delta v}{2\sqrt{\zeta b}}.
\end{equation}
Under cyberattack conditions, the measured spacing $\tilde{s}$, speed $\tilde{v}$, and relative speed $\Delta \tilde{v}$ are modeled as follows:
\begin{equation}
\begin{aligned}
\tilde{s} &= s+\alpha_s,\\
\tilde{v} &= v+\alpha_v,\\
\Delta \tilde{v} &= \Delta v+\alpha_{\Delta v},
\end{aligned}
\label{eq:measurement_attack}
\end{equation}
where \(\alpha_s\), \(\alpha_v\), and \(\alpha_{\Delta v}\) denote attack-induced corruptions to the spacing, speed, and relative speed measurements, respectively. 
Subsequently, the acceleration computed from corrupted measurements is given by:
\begin{equation}
\hat{a}
=
F\!\left(
\boldsymbol{\theta},
\tilde{s},
\tilde{v},
 \Delta \tilde{v}
\right).
\label{eq:attacked_acceleration}
\end{equation}
For local sensitivity analysis, a first-order Taylor expansion around the nominal operating point 
\((s,v,\Delta v)\) gives
\begin{equation}
\begin{aligned}
\hat{a}
&=
a+d +
\mathcal{O}\!\left(
\alpha_s^2+
\alpha_v^2+
\alpha_{\Delta v}^2
\right),
\end{aligned}
\label{eq:taylor_expansion}
\end{equation}
where \(a=F(\boldsymbol{\theta},s,v,\Delta v)\), and
\begin{equation}
d =
\frac{\partial F}{\partial s}\alpha_s
+
\frac{\partial F}{\partial v}\alpha_v
+
\frac{\partial F}{\partial \Delta v}\alpha_{\Delta v}.
\label{eq:disturbance}
\end{equation}

The partial derivatives are evaluated at the nominal operating point \((s,v,\Delta v)\). 
The Taylor-expanded form is used only to interpret the local effect of measurement corruption on IDM acceleration. 

Let $\mathcal{D}^{E}$ denote the attack-free expert trajectories generated by the IDM calibrated using a dataset. The steady-state calibration subset is $\mathcal{D}_{0}=\left\{O_{t}^{E}\in \mathcal{D}^{E}\mid t>t_{\mathrm{ss}}\right\}$, where $O_{t}^{E}$ denotes the nominal expert state and $t_{\mathrm{ss}}$ is the warm-up threshold.
The RAT mechanism directly computes the RDC variables from the observed driving state using the exact IDM sensitivities, without relying on the first-order approximation. For the IDM, these sensitivities are given by:
\begin{equation}
\begin{aligned}
&\frac{\partial F}{\partial s}
=
2\zeta
\frac{(s^*)^2}{s^3},\\[1ex]
&\frac{\partial F}{\partial \Delta v}
=
2\zeta
\frac{s^*}{s^2}
\frac{v}{2\sqrt{\zeta b}},\\[1ex]
&\frac{\partial F}{\partial v}
=
-\zeta\delta
\frac{v^{\delta-1}}{v_0^\delta}
-
2\zeta
\frac{s^*}{s^2}
\left(
T-\frac{\Delta v}{2\sqrt{\zeta b}}
\right).
\end{aligned}
\label{eq:idm_sensitivity}
\end{equation}
These derivatives characterize the sensitivity of IDM acceleration to spacing, relative speed, and speed. Accordingly, the RDC variables are defined as:

\begin{equation}
\beta_1=\frac{\partial F}{\partial s},\quad
\beta_2=\frac{\partial F}{\partial \Delta v},\quad
\beta_3=\frac{\partial F}{\partial v}.
\label{eq:rdc_variables}
\end{equation}


Following established rationality conditions for car-following dynamics \cite{wang2024analytical,wang2024string}, the IDM sensitivities are required to satisfy:
\begin{equation}
\left\{
\begin{aligned}
\beta_1 &> 0,\\
\beta_2 &> 0,\\
\beta_3 &< 0.
\end{aligned}
\right.
\label{eq:beta}
\end{equation}
The admissible RDC intervals, which are subsequently enforced as fixed bounds, are calibrated as follows:
\begin{equation}
\mathcal{C}_k
=
[\beta_k^{-},\beta_k^{+}],
\quad k=1,2,3,
\end{equation}
where
\begin{equation}
\beta_k^{-}
=
\min_{O^E_{\tau}\in\mathcal{D}_0}
\beta_k(O^E_{\tau}),
\quad
\beta_k^{+}
=
\max_{O^E_{\tau}\in\mathcal{D}_0}
\beta_k(O^E_{\tau}).
\end{equation}
Here, \(\beta_k(O^E_{\tau})\) denotes the \(k\)-th RDC variable computed from attack-free trajectories at time step \(\tau\). 
The admissible RDC set is defined as the Cartesian product:
\begin{equation}
\mathcal{C}
=
\mathcal{C}_1
\times
\mathcal{C}_2
\times
\mathcal{C}_3.
\label{eq:rdc_set}
\end{equation}
These calibrated intervals are fixed after calibration and serve as hard rationality boundaries for accurately identifying RDC violations and triggering CRIL-based trajectory recovery. We denote the above calibration procedures by $\operatorname{CalibrateRDC}
\!\left(
\mathcal{D}_0, \boldsymbol{\theta}
\right)$.

RAT computes the RDC variables from the observed corrupted state
$(\tilde{s}, \tilde{v}, \Delta\tilde{v})$, denoted by
$\tilde{\beta}_1$, $\tilde{\beta}_2$, and $\tilde{\beta}_3$. The observed corrupted state is considered admissible if
\begin{equation}
\left\{
\begin{aligned}
&0 < \beta_1^{-}
\leq \tilde{\beta}_1
\leq \beta_1^{+},\\
&0 < \beta_2^{-}
\leq \tilde{\beta}_2
\leq \beta_2^{+},\\
&\beta_3^{-}
\leq \tilde{\beta}_3
\leq \beta_3^{+}<0.
\end{aligned}
\right.
\label{eq:hard_rdc}
\end{equation}
Let the RDC vector be
$\tilde{\boldsymbol{\beta}}(t)
=
[\tilde{\beta}_1(t),\tilde{\beta}_2(t),\tilde{\beta}_3(t)]^\top$.
The RAT triggering signal is given by:
\begin{equation}
\Gamma(t)
=
\mathbb{I}\!\left[
\tilde{\boldsymbol{\beta}}(t)\notin \mathcal{C}
\right],
\label{eq:trigger}
\end{equation}
where $\mathbb{I}[\cdot]$ denotes the indicator function, with
$\Gamma(t)=1$ indicating an RDC violation and the activation of CRIL, and $\Gamma(t)=0$ otherwise.
Thus, recovery is triggered when at least one RDC variable falls outside
its calibrated admissible interval, $i.e.,$ there exists $k\in\{1,2,3\}$ such that

\begin{equation}
\tilde{\beta}_k<\beta_k^{-}
\quad
\text{or}
\quad
\tilde{\beta}_k>\beta_k^{+}.
\label{eq:rdc_violation}
\end{equation}

\subsection{Cross-Branch Robust Imitation Learning}

Upon RAT activation, CRIL reconstructs clean states from corrupted observations and generates recovery actions through the learned policy.
At each time step $t$, the cyberattack-corrupted observation is represented as:
\begin{equation}
O_t^{obs}
=
[\tilde{s}(t),\tilde{v}(t),\Delta \tilde{v}(t)]^{\top}.
\end{equation}
The nominal expert state from attack-free trajectories $\mathcal{D}_E$ and its corresponding action are defined as:
\begin{equation}
O_t^E
=
\left[s^E(t),v^E(t),\Delta v^E(t)\right]^\top,
\qquad
u_t^E=v^E(t+1),
\label{eq:expert_demonstration}
\end{equation}
where $u_t^E$ denotes the one-step-ahead expert speed associated with
$O_t^E$. 

During training, the state generator $G_{\psi}$ is optimized as a
denoising reconstructor. Measurement perturbations are injected into attack-free expert trajectories to construct a perturbed
historical window:
\begin{equation}
H_t^{p}
=
\bigl[
O_{t-W+1}^{E}+\epsilon_{t-W+1}^{G},
\ldots,
O_t^{E}+\epsilon_t^{G}
\bigr],
\;
\epsilon_{\tau}^{G}
\sim
\mathcal{N}\!\left(0,\sigma_G^{2}I\right),
\label{eq:perturbed_window}
\end{equation}
where $W$ denotes the historical window length. 
%
The operator $\operatorname{DenoisingTrain}(\cdot)$ updates $G_{\psi}$  by minimizing $\mathcal{L}_G$. After convergence, the reconstructed nominal state is obtained as:
\begin{equation}\label{eq:generator}
\bar{O}_t
=
G_{\psi}\left(H_t^{p}\right),
\end{equation}
and
\begin{equation}\label{eq:generator_loss}
\mathcal{L}_{G}
=
\mathbb{E}
\left[
\left\|
\bar{O}_t-O_t^{E}
\right\|_{2}^{2}
\right].
\end{equation}
This denoising objective enables $G_{\psi}$ to map perturbed historical observations toward the nominal driving-state domain.

\begin{algorithm}
\caption{Training of TRACER}
\label{alg:TRACER_training}
\begin{algorithmic}[1]

\Statex \textbf{Input:} ICE-ACC dataset
\Statex \textbf{Output:} IDM parameter $\boldsymbol{\theta}$, RDC region $\mathcal{C}$, state generator $G_{\psi}^{*}$, 
and recovery policy $\pi^{*}$
\State $\boldsymbol{\theta} \leftarrow$ Calibrate IDM parameters by Eqs. (\ref{eq:acceleration})-(\ref{eq:desired_spacing})
\State $\mathcal{D}_E \leftarrow \text{Simulate attack-free trajectories using } \boldsymbol{\theta}$
\State $\mathcal{D}_0 \leftarrow \left\{ \mathcal{D}_t^E \in \mathcal{D}^E \mid t > t_{ss} \right\}$
\State $\mathcal{C} \gets \operatorname{CalibrateRDC}(\mathcal{D}_{0},\boldsymbol{\theta})$ by Eqs. (\ref{eq:idm_sensitivity})-(\ref{eq:rdc_set})
\State $H_t^{p} \gets $ Inject perturbations to
$O_{t-W+1:t}^{E}$

\Repeat
    \State $G_{\psi}
    \gets \operatorname{DenoisingTrain}
    (H_t^{p},O_t^{E})$
\Until{$G_{\psi}$ converges}
\State $G_{\psi}^{*} \gets G_{\psi}$
\State $\bar{O}_t \leftarrow G_{\psi}^{*}\left(H_t^{p}\right)$

\Repeat
    \State $(\pi_{\phi_1},D_{1})
    \gets \operatorname{AIL}\!\left(
    \{(O_t^{E},u_t^{E})\},
    \{(\bar{O}_t,
    \pi_{\phi_1}(\bar{O}_t))\}
    \right)$
\Until{$\pi_{\phi_1}$ and $D_{1}$ converge}

\State $\hat{O}_t^{E}
\gets O_t^{E}+\epsilon_{E}$; $\hat{O}_t
\gets \bar{O}_t+\epsilon_{R}$

\Repeat
    \State $(\pi_{\phi_2},D_{2})
    \gets \operatorname{AIL}\!\left(
    \{(\hat{O}_t^{E},u_t^{E})\},
    \{(\hat{O}_t,
    \pi_{\phi_2}(\hat{O}_t))\}
    \right)$
\Until{$\pi_{\phi_2}$ and $D_{2}$ converge}

\State Freeze $D_{1}$

\Repeat
    \State $\pi_{\phi_2}
    \gets \operatorname{CrossBranchFT}\!\left(
    \{(\bar{O}_t,
    \pi_{\phi_2}(\bar{O}_t))\};D_{1}
    \right)$
\Until{$\pi_{\phi_2}$ converges}

\State $\pi^{*} \gets \pi_{\phi^*_2}$

\end{algorithmic}
\end{algorithm}

\begin{algorithm}
\caption{Recovery of TRACER}
\label{alg:TRACER_testing}
\begin{algorithmic}[1]

\Statex \textbf{Input:} Corrupted observations $O_t^{obs}$, IDM parameter $\boldsymbol{\theta}$, RDC region $\mathcal{C}$, state generator $G_{\psi}^{*}$, 
and recovery policy $\pi^{*}$
\Statex \textbf{Output:} Recovered trajectories $O^r_{t+1}$
\For{$t$}

    \State Adopt $\boldsymbol{\theta}$ to compute $\tilde{\boldsymbol{\beta}}(t)$ by Eqs. (\ref{eq:idm_sensitivity})-(\ref{eq:rdc_variables})
    \State $\Gamma(t) \leftarrow \mathbb{I}\!\left[\tilde{\boldsymbol{\beta}}(t) \notin \mathcal{C}\right]$
    \If{$\Gamma(t) = 1$}
    \State $\bar{O}_t \leftarrow G_{\psi}^{*}(H_t^{obs})$
        \State 
        $v^r(t+1)=\pi^{*}(\bar{O}_t)$

        \State Update $O^r_{t+1}$ by Eqs. (\ref{eq:recovered_relative_speed})-(\ref{eq:recovered_state})

    \Else

        \State Continue normal ACC control

    \EndIf

\EndFor

\end{algorithmic}
\end{algorithm}

We construct two adversarial policy–discriminator branches to preserve behavioral fidelity and enhance robustness to observation perturbations. In each branch, the policy and discriminator are alternately optimized through an adversarial imitation learning operator, denoted by $\operatorname{AIL}(\cdot)$.
The clean branch learns nominal car-following behavior using expert state--action pairs and reconstructed clean states:
\begin{equation}
\begin{split}
\min_{\phi_1}\max_{D_1}\; 
&
\mathbb{E}_{(O_t^E,u_t^E)\sim\rho_E}
\left[
\log D_1
\left(
O_t^E,
u_t^E
\right)
\right] \\
&+
\mathbb{E}_{\bar{O}_t\sim\rho_{\pi_{\phi_1}}}
\left[
\log
\left(
1-
D_1
\left(
\bar{O}_t,
\pi_{\phi_1}(\bar{O}_t)
\right)
\right)
\right],
\end{split}
\label{eq:D1}
\end{equation}
where \(\pi_{\phi_1}\) denotes the clean-branch policy parameterized by \(\phi_1\), \(D_1\) is the clean-branch discriminator, and \(\rho_{\pi_{\phi_1}}\) and $\rho_E$ denote the occupancy distribution induced by \(\pi_{\phi_1}\) and the expert policy, respectively. 

To enhance robustness to observation perturbations, the noisy branch introduces Gaussian noise into the expert and reconstructed states as follows:
\begin{equation}
\hat{O}_t^E
=
O_t^E+\epsilon_E,
\quad
\hat{O}_t
=
\bar{O}_t+\epsilon_R,
\quad
\epsilon_E,\epsilon_R
\sim
\mathcal{N}(0,\sigma^2 I).
\label{eq:noisy_state}
\end{equation}
The noisy branch is optimized by:
\begin{equation}
\begin{split}
\min_{\phi_2}\max_{D_2}\; 
&
\mathbb{E}_{(\hat{O}_t^E,u_t^E)\sim\hat{\rho}_E}
\left[
\log D_2
\left(
\hat{O}_t^E,
u_t^E
\right)
\right] \\
&+
\mathbb{E}_{\hat{O}_t\sim\rho_{\pi_{\phi_2}}}
\left[
\log
\left(
1-
D_2
\left(
\hat{O}_t,
\pi_{\phi_2}(\hat{O}_t)
\right)
\right)
\right].
\end{split}
\label{eq:D2}
\end{equation}
The clean branch captures nominal driving characteristics, while the noisy branch improves policy robustness to observation perturbations. We further define a cross-branch fine-tuning operator, denoted by $\operatorname{CrossBranchFT}(\cdot)$, which updates the noisy-branch policy using the fixed clean-branch discriminator $D_1$ by:
\begin{equation}
\phi_2^{*}
=
\arg\min_{\phi_{2}}
\left\{
-\mathbb{E}_{\bar{O}_{t}\sim \rho_{\pi_{\phi_2}}}
\left[
\log D_{1}
\left(
\bar{O}_{t},
\pi_{\phi_{2}}(\bar{O}_{t})
\right)
\right]
\right\}.
\label{fine-tuning}
\end{equation}
This objective regularizes the robust policy toward the nominal driving distribution while retaining the robustness acquired from noisy-branch training. 
The refined noisy-branch policy is therefore adopted as the final recovery policy, $\textit{i.e.}$, $\pi^{*} = \pi_{\phi_{2}^{*}}$.

During recovery, the corrupted historical window is constructed as
\begin{equation}
H_t^{obs}
=
\left[
O_{t-W+1}^{\mathrm{obs}},
\ldots,
O_t^{\mathrm{obs}}
\right],
\end{equation}
and the trained generator reconstructs the nominal state as
\begin{equation}
\bar{O}_t
=
G_{\psi}^{*}\left(H_t^{\mathrm{obs}}\right).
\end{equation}
The final recovery policy generates the recovered velocity of vehicle $i$ as:
\begin{equation}
v^r(t+1)=\pi^{*}(\bar{O}_t).
\end{equation}
Based on the state definitions in Eqs. (\ref{eq:spacing}) to (\ref{eq:relative speed}), the recovered relative speed and spacing are updated as:
\begin{equation}\label{eq:recovered_relative_speed}
\Delta v^r(t+1)=v_{i-1}(t+1)-v^r(t+1),
\end{equation}
\begin{equation}\label{eq:recovered_spacing}
s^r(t+1)=s^r(t)+\Delta t\,\Delta v^r(t+1),
\end{equation}
where $\Delta t$ denotes the sampling interval. Hence, the recovered state can be obtained as:
\begin{equation}\label{eq:recovered_state}
O_{t+1}^r=
\left[
s^r(t+1),
v^r(t+1),
\Delta v^r(t+1)
\right]^{\top}.
\end{equation}

\section{Experiments}\label{experiment}
This section presents a comprehensive evaluation of the proposed TRACER framework under diverse cyberattack scenarios, most of which are communication-based cyberattacks.


\subsection{Experimental setup}\label{setup}

\begin{table*}[!t]
\centering
\caption{Summary of adopted cyberattack scenarios}
\label{tab:attack_summary_risk}
\renewcommand{\arraystretch}{1.15}
\small

\begin{tabularx}{\textwidth}{
>{\centering\arraybackslash}m{0.9cm}
>{\centering\arraybackslash}m{1.0cm}
>{\raggedright\arraybackslash}X
>{\centering\arraybackslash}m{1.0cm}
>{\centering\arraybackslash}m{1.2cm}
}
\hline

\noalign{\vskip 2pt}
Attack & Risk & \multicolumn{1}{c}{Description} & $\tilde{R}_v$ & $\Delta \tilde{R}_v$ \\
\hline

MaA 
& \multirow{3}{*}{High}
& Perturbs the ACC acceleration command with an additive disturbance
& 19.949 & 19.452 \\

BA  
&
& Redirects the ACC vehicle to a more distant vehicle by obscuring intermediate vehicles
& 13.724 & 13.227 \\

DPA 
&
& Perturbs the ACC spacing and relative-speed measurements with injected disturbances
& 2.959 & 2.462 \\

\hline

PA  
& Medium
& Replaces the vehicle's leader-related information with that associated with another vehicle
& 1.010 & 0.513 \\

\hline

FA  
& \multirow{2}{*}{Low}
& Fixes the ACC speed input at its value at the attack onset
& 0.010 & 0.487 \\

MA  
&
& Combines mismatched leader-related information with a discrete communication delay
& 0.808 & 0.311 \\

\hline
\end{tabularx}
\end{table*}

\noindent \textbf{Datasets}. Representative traffic datasets for microscopic car-following modeling include NGSIM \cite{coifman2017critical}, HighD \cite{krajewski2018highd}, I-24 MOTION \cite{gloudemans202324}, and ICE-ACC \cite{zhou2024review}. NGSIM contains vehicle trajectories collected on U.S. freeways but does not explicitly identify production ACC vehicles. HighD provides high-resolution trajectories extracted from drone recordings, although continuous observations of individual vehicles are relatively limited. I-24 MOTION offers large-scale freeway trajectory data but requires additional leader--follower reconstruction for car-following analysis. ICE-ACC directly records longitudinal states of production ACC-equipped vehicles, including speed, spacing, relative speed, and acceleration, making it well aligned with the state representation adopted in this study. Therefore, ICE-ACC is selected to calibrate the ACC vehicle model.

\noindent \textbf{Cyberattack scenarios}. 
%
This study considers six representative cyberattack scenarios, as summarized in Table \ref{tab:attack_summary_risk}. Four communication-based attacks from \cite{li2025conceptualizing}, including fixed speed attack (FA), phantom attack (PA), blinding attack (BA), and mixed attack (MA), and two additional attacks from \cite{li2024detecting}, namely malicious acceleration attack (MaA) and data poisoning attack (DPA), are adopted. 
Attack severity is classified into three risk levels based on 
$\Delta\tilde{R}_v = |\tilde{R}_v - R_v|$, using thresholds of $0.5$ and $1$: 
low risk for $\Delta\tilde{R}_v < 0.5$, medium risk for 
$0.5 \leq \Delta\tilde{R}_v < 1$, and high risk for 
$\Delta\tilde{R}_v \geq 1$, where $\tilde{R}_v$ denotes the attacked speed fluctuation amplification ratio and $R_v = 0.497$ is the attack-free reference. The resulting classification reflects the degree to which each attack disrupts traffic stability and car-following dynamics.

Under MaA, an additive disturbance $\xi$ is directly injected into the ACC acceleration command:
\begin{equation}
\hat{a}_i(t)
=
{\Pi}_{[g,h]}
\left[
F(\boldsymbol{\theta}, s, v, \Delta v)+\xi
\right],
\label{MaA}
\end{equation}
\noindent where $\xi\sim\mathcal{N}(0,5)$ and $\Pi_{[g,h]}(\cdot)$ bounds the resulting acceleration within $[g,h]$.

Under BA, the vehicle fails to perceive intermediate vehicles and responds to a more distant vehicle $j$ as its leader:
\begin{equation}
\hat{a}_i(t)
=
F\left(\boldsymbol{\theta}, \tilde{s}, v_i, v_j-v_i\right),
\label{BA}
\end{equation}
\noindent where $\tilde{s}$ denotes the cumulative perceived spacing from vehicle $i$ to the distant vehicle $j$.
Under DPA, the spacing and relative-speed measurements are perturbed before being used by the ACC controller:
\begin{equation}
\hat{a}_i(t)
=
\Pi_{[g,h]}
F\left(\boldsymbol{\theta}, s+\lambda_1, v, \Delta v+\lambda_2\right),
\label{DPA}
\end{equation}
\noindent 
where $\lambda_1$ and $\lambda_2$ are independent perturbations satisfying $\lambda_1,\lambda_2\sim\mathcal{N}(0,5)$.

PA is classified as a medium-risk attack. Under PA, the vehicle receives the leader-related information of another vehicle $j$ instead of that associated with its own leader:
\begin{equation}
    \hat{a}_i(t) = F(\boldsymbol{\theta}, s_j, v, v_{j-1}-v_j),
    \label{PA}
\end{equation}
\noindent where $s_j$ and $v_{j-1}-v_j$ denote the spoofed spacing and relative speed, respectively, while $v$ is the speed of the attacked vehicle.

FA and MA are categorized as low-risk attacks. Under FA, the ACC uses the speed measured at the attack onset throughout the attack:
\begin{equation}
    \hat{a}_i(t)= F(\boldsymbol{\theta},s,  v(\tilde{t}),  \Delta v  ),
    \label{FA}
\end{equation}
\noindent where $\tilde{t}$ denotes the attack onset time.
%
%
Under MA, mismatched leader-related information is combined with a communication delay:
\begin{align}
\hat{a}_i(t)
&= F\Big(
\boldsymbol{\theta},
s_j(\lfloor t-m\rfloor),
v(t),
\notag\\
&\quad
v_{j-1}(\lfloor t-m\rfloor)
-
v_j(\lfloor t-m\rfloor)
\Big),
\label{MA}
\end{align}
\noindent where $m$ denotes the communication delay in discrete simulation steps. Thus, the attacked vehicle uses delayed spacing and relative-speed information from vehicle $j$ instead of its own current leader-related information.

\noindent \textbf{Baselines for comparison}. 
To comprehensively evaluate the effectiveness of the proposed method, six representative trajectory recovery approaches were selected as baselines. Transformer \cite{zhu2022transfollower} and LSTM \cite{ko2021approach} were employed as sequence modeling methods for trajectory reconstruction from historical observations. A diffusion model \cite{you2025followgen} was included as a representative generative approach based on iterative denoising. GAIL \cite{ho2016generative}, H-GAIL \cite{couto2024hierarchical}, and DRAIL \cite{lai2024diffusion} were selected as representative imitation learning methods, where H-GAIL employs hierarchical policy learning and DRAIL incorporates diffusion-based reward modeling for enhanced robustness. These baselines cover sequence modeling, generative modeling, and imitation learning paradigms, enabling a comprehensive comparison with the proposed CRIL.

\noindent \textbf{Implementation details}. The IDM parameters for the ACC-equipped vehicle were calibrated using ICE-ACC trajectories to characterize its longitudinal car-following dynamics. The calibrated parameter values are reported in Table \ref{tab:idm_parameters}. For the human-driven vehicles (HDVs), the IDM parameters were adopted from the calibrated settings reported in \cite{kesting2008calibrating}.

\begin{table}[htbp]
\centering
\caption{IDM parameter settings for ACC and human-driven vehicles}
\label{tab:idm_parameters}

\resizebox{\columnwidth}{!}{%
\begin{tabular}{ccccccc}
\hline
\noalign{\vskip 2pt}
\multirow{2}{*}[-0.8pt]{Vehicle Type}
& $\zeta$
& $b$
& $\delta$
& $s_0$
& $T$
& $v_0$ \\
\noalign{\vskip 1pt}
\cline{2-7}
\noalign{\vskip 2pt}
& $(m/s^2)$
& $(m/s^2)$
& $-$
& $(m)$
& $(s)$
& $(m/s)$ \\
\noalign{\vskip 1pt}
\hline
\noalign{\vskip 2pt}
\multirow{1}{*}{ACC}
& 1.18
& 8.90
& 20.00
& 10.00
& 0.96
& 39.12 \\

\multirow{1}{*}{HDV}
& 1.06
& 2.00
& 4.00
& 3.40
& 1.26
& 30.00 \\

\hline

\end{tabular}
}
\end{table}

\begin{figure}[h]
\centering
\includegraphics[width=0.6\columnwidth]{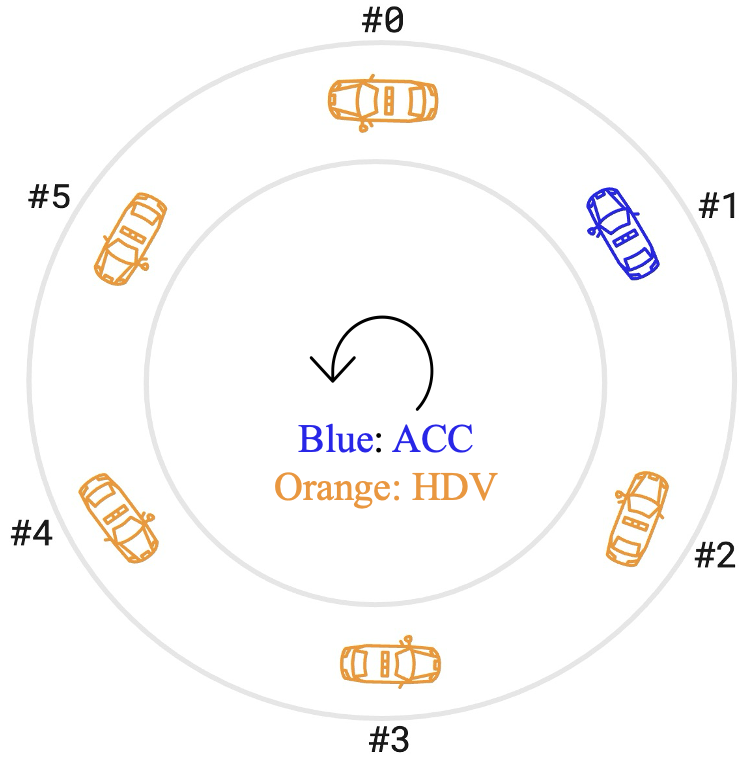}
\caption{
Ring-road simulation with six vehicles in mixed traffic, the ACC vehicle (blue) is subjected to a cyberattack.
}
\label{fig:simulation}
\end{figure}

Based on the calibrated vehicle models, a microscopic traffic simulation was conducted on a 300-meter single-lane ring road, a widely used setting for car-following and traffic-flow studies \cite{li2024can}, as shown in Fig. \ref{fig:simulation}. The platoon contained six vehicles, including one ACC-equipped vehicle and five HDVs. Each simulation lasted eight minutes, with a 30-second cyberattack imposed on the ACC vehicle over the attack interval $\mathcal{T}={t_1,t_2,\ldots,t_K}$, where $K=300$ time steps. Vehicle states were sampled at 10 Hz to obtain speed, spacing, and relative-speed trajectories for evaluation. For RDC calibration, $t_{ss}=2000$ was used to exclude initial transient dynamics.

To ensure a fair and reproducible comparison, all methods were evaluated using the same data partition, with the training, validation, and testing sets split in a 70:15:15 ratio.
Each method was independently evaluated over five runs under identical experimental conditions. The key hyperparameters for all baseline methods are summarized in Table \ref{tab:hyperparameters}.

\begin{table}[htbp]
\centering
\caption{Hyperparameter settings for baseline methods}
\label{tab:hyperparameters}

\scriptsize
\renewcommand{\arraystretch}{1.25}
\setlength{\tabcolsep}{2.0pt}

\begin{tabular}{lccccccc}
\hline
Parameter
& GAIL
& H-GAIL
& DRAIL
& LSTM
& Transformer
& Diffusion
& CRIL \\

\hline

Batch size
& 64 & 64 & 64 & 64 & 64 & 64 & 64 \\

Learning rate
& 1e-3 & 1e-3 & 1e-3 & 1e-4 & 1e-4 & 1e-4 & 1e-3 \\

Training epochs
& 60 & 60 & 60 & 60 & 60 & 60 & 60 \\

Generator epochs
& - & 60 & - & - & - & - & 60 \\


Window size
& - & 8 & - & - & - & - & 8 \\

Sequence length
& - & - & - & 10 & 10 & 10 & - \\

Hidden dim.
& - & - & - & 128 & 128 & 128 & - \\

\hline
\end{tabular}

\renewcommand{\arraystretch}{1}
\setlength{\tabcolsep}{6pt}

\end{table}

CRIL employs a Bi-LSTM generator with fully connected layers to reconstruct nominal states from historical observations. The clean and noise-enhanced policies are implemented as separate two-hidden-layer MLPs with ReLU activations. Their corresponding discriminators are also independently parameterized MLPs, using ReLU activations in the hidden layers and Sigmoid output activations.

\noindent \textbf{Evaluation metrics}. 

Trajectory recovery performance is evaluated from three aspects: recovery accuracy, traffic stability, and safety preservation. These metrics assess the consistency between recovered and nominal trajectories, as well as the ability of the recovered trajectories to maintain physically plausible car-following behavior and safe inter-vehicle spacing under cyberattacks.

Let $v^r(t)$ and $v^E(t)$ denote the recovered and attack-free reference velocities over the attack interval $\mathcal{T}$, respectively, where $K=|\mathcal{T}|$. The velocity recovery accuracy is evaluated using root mean square error (RMSE), mean absolute error (MAE), and mean absolute percentage error (MAPE):

\begin{equation}
\text{RMSE}_v =
\sqrt{\frac{1}{K}\sum_{t\in\mathcal{T}}\left(v^r(t)-v^E(t)\right)^2}
\end{equation}

\begin{equation}
\text{MAE}_v =
\frac{1}{K}\sum_{t\in\mathcal{T}}\left|v^r(t)-v^E(t)\right|
\end{equation}

\begin{equation}
\text{MAPE}_v =
\frac{100 \%}{K}\sum_{t\in\mathcal{T}}\left|\frac{v^r(t)-v^E(t)}{v^E(t)}\right|
\end{equation}
Accordingly, the spacing recovery metrics $\text{RMSE}_s$, $\text{MAE}_s$, and $\text{MAPE}_s$ are computed in a similar manner.

Inspired by the characterization of traffic oscillation propagation using vehicle speed standard deviation in \cite{jiang2018experimental}, this study defines the speed fluctuation amplification ratio, $R_v$, to quantify the propagation and amplification of speed disturbances between consecutive vehicles in a platoon by:
\begin{equation}
R_v
=
\frac{
\sqrt{
\frac{1}{K}
\sum_{t\in\mathcal{T}}
\left(
v^E_i(t)-\bar{v}^E_i
\right)^2
}
}{
\sqrt{
\frac{1}{K}
\sum_{t\in\mathcal{T}}
\left(
v_{i-1}(t)-\bar{v}_{i-1}
\right)^2
}
},
\end{equation}
where
\begin{equation}
\bar{v}^E_i
=
\frac{1}{K}
\sum_{t\in\mathcal{T}}
v^E_i(t),
\qquad
\bar{v}_{i-1}
=
\frac{1}{K}
\sum_{t\in\mathcal{T}}
v_{i-1}(t).
\end{equation}
The recovered and attacked counterparts, denoted by $R_v^r$ and $\tilde{R}_v$, are computed using the same denominator, with the numerator calculated from $v_i^r(t)$ and $\tilde{v}_i(t)$, respectively.
The deviations from the attack-free reference behavior are then defined as:
\begin{equation}
\Delta R_v^r
=
\left|
R_v^r-R_v
\right|,
\qquad
\Delta \tilde{R}_v
=
\left|
\tilde{R}_v-R_v
\right|.
\end{equation}
A smaller $\Delta R_v^r$ indicates that the recovered trajectory better preserves the attack-free car-following behavior.

Safety preservation is evaluated using the minimum spacing over the attack interval $\mathcal{T}$. The minimum spacings of the attack-free, recovered, and attacked trajectories are computed as:
\begin{equation}
\begin{aligned}
s^E_{\min}
&=
\min_{t\in\mathcal{T}} s^E(t),
&
s_{\min}^{r}
&=
\min_{t\in\mathcal{T}} s^{r}(t),
&
\tilde{s}_{\min}
&=
\min_{t\in\mathcal{T}} \tilde{s}(t).
\end{aligned}
\end{equation}
The corresponding deviations from the attack-free reference are defined as:
\begin{equation}
\begin{aligned}
\Delta s_{\min}^{r}
&=
\left|
s_{\min}^{r}
-
s^E_{\min}
\right|,
&
\Delta \tilde{s}_{\min}
&=
\left|
\tilde{s}_{\min}
-
s^E_{\min}
\right|.
\end{aligned}
\end{equation}
A smaller $\Delta s_{\min}^{r}$ indicates better preservation of the attack-free safety margin.

\subsection{RAT Triggering Performance Evaluation}

To evaluate the reliability and robustness of RAT, ten attack episodes ($N_{\mathrm{attack}}=10$) were conducted for each cyberattack scenario. For each episode, the attack interval was randomly selected from 3400–3700, 3500–3800, and 3600–3900 time steps, while the numbers of ACC-equipped and attacked ACC vehicles were varied between one and two. 
Accordingly, each episode was generated by randomly sampling an attack interval and a feasible ACC configuration. The same experimental procedure was applied to all six cyberattack scenarios to evaluate RAT under different attack intervals, ACC deployment levels, and attack scales.

Triggering performance was evaluated using the trigger success rate
($TSR$) and triggering delay $T_{\mathrm{delay}}$. The $TSR$ is
defined as:
\begin{equation}
TSR
=
\frac{N_{\mathrm{trigger}}}{N_{\mathrm{attack}}}
\times 100\%,
\end{equation}
where $N_{\mathrm{trigger}}$ denotes the number of successfully triggered episodes. The triggering delay is calculated as:
\begin{equation}
T_{\mathrm{delay}}
=
t_{\mathrm{trigger}}-t_{\mathrm{attack}},
\end{equation}
where $t_{\mathrm{attack}}$ and $t_{\mathrm{trigger}}$ denote the
attack onset and corresponding triggering times, respectively. A
smaller $T_{\mathrm{delay}}$ indicates a faster response to the attack.

\begin{table}[h]
\centering
\caption{RAT performance in abnormal-state triggering across ten experiments for each cyberattack scenario
}
\label{tab:RAT}
\renewcommand{\arraystretch}{1.15}

\begin{tabular*}{\linewidth}{@{\extracolsep{\fill}}ccc}
\hline
Attack &
\textbf{$TSR$} $\uparrow$ &
$T_{\mathrm{delay}} \downarrow$ \\
\midrule
FA & 100\% & 8.68 $\pm$ 4.02\\
PA & 100\% & $<$ 0.1s \\
BA & 100\% & $<$ 0.1s \\
MA & 100\% & $<$ 0.1s \\
MaA &100\% & $<$ 0.1s \\
DPA & 100\% &$<$ 0.1s \\
\hline
\end{tabular*}

\end{table}

As summarized in Table \ref{tab:RAT}, RAT successfully triggered recovery in all evaluated attack episodes, achieving a trigger success rate of 100$\%$ in each case. These results demonstrate the robustness and generalizability of the proposed triggering mechanism under varying traffic and attack conditions.

Notably, FA yields an average triggering delay of $8.68 \pm 4.02$ seconds since it forces the attacked vehicle to maintain its velocity at attack onset. Under stable car-following, the resulting deviation is initially small and does not immediately violate the RDC bounds. As the attack persists, the velocity mismatch accumulates into larger spacing errors, eventually driving the RDC variables outside their admissible ranges and activating RAT. Hence, the observed delay reflects the gradual propagation of FA-induced disturbances, with minor variations arising from differences in attack timing and traffic conditions.

\begin{table*}[!t]
\caption{
Trajectory recovery performance of different methods. The best results are highlighted in bold, and the second-best results are underlined
}
\label{Numerical comparison of different methods}
\resizebox{\linewidth}{!}{
\begin{tabular}{>{\centering\arraybackslash}p{0.3cm} >{\centering\arraybackslash}p{2cm} |cccccc|cccc}
\hline
\multicolumn{2}{c|}{} 
& \multicolumn{3}{c}{Spacing} 
& \multicolumn{3}{c|}{Speed} 
& \multicolumn{2}{c}{Post-defense} 
& \multicolumn{2}{c}{Deviation} \\ \cline{3-12}

\multirow{2}{*}[2.5 ex]{Attack} 
& \multirow{2}{*}[2.5 ex]{Method} 
& $\text{MAE}_s$ $\downarrow$ & $\text{RMSE}_s \downarrow$ &  $\text{MAPE}_s  \downarrow$ 
&  $\text{MAE}_v \downarrow$ &  $\text{RMSE}_v  \downarrow$ &  $\text{MAPE}_v  \downarrow$
& $R_v^{r}$ & $s_{\min}^{r}$ 
& $\Delta R_v^{}$ $\downarrow$ & $\Delta s_{\min}^r$ $\downarrow$ \\ 
\hline

\multirow{7}{*}{FA}    
& H-GAIL  & 0.453$\pm$0.190& 0.546$\pm$0.193& 1.35\%& 0.134$\pm$0.004& 0.162$\pm$0.006& 0.54\%&  0.789&  33.626&  0.292&  \underline{0.215}\\
& DRAIL   & 0.503$\pm$0.256& 0.556$\pm$0.259& 1.49\%& 0.137$\pm$0.011& 0.173$\pm$0.008& 0.56\%&  0.976&  32.937&  0.479&  0.474\\
& GAIL    & \underline{0.345$\pm$0.112}& \underline{0.416$\pm$0.116}& \underline{1.02\%}& 0.139$\pm$0.010& 0.227$\pm$0.009& 0.56\%&  0.925&  33.953&  0.428&  0.542\\

& Transformer    & 10.039$\pm$0.075& 10.058$\pm$0.075& 29.78\%& \textbf{0.060$\pm$0.001}& \textbf{0.075$\pm$0.001}& \textbf{0.24\%}&  0.048&  43.139&  0.449&  9.728\\
& LSTM    & 1.979$\pm$0.236 & 2.472$\pm$0.283& 5.88\%& 0.136$\pm$0.112& 0.222$\pm$0.252& 0.55\%&  1.068&  33.202&  0.571 & 0.209 \\
& Diffusion    & 0.530$\pm$0.042 & 0.682$\pm$0.046 & 1.57\%& \underline{0.096$\pm$0.004}& \underline{0.119$\pm$0.007}& \underline{0.39\%}&  0.213&  33.058& \underline{0.284}&  0.353\\

& CRIL & \textbf{0.317$\pm$0.070}& \textbf{0.391$\pm$0.081}& \textbf{0.94\%}& 0.121$\pm$0.001& 0.148$\pm$0.002& 0.51\%&  0.776&  33.294&  \textbf{0.279}&  \textbf{0.117}\\ 
\hline

\multirow{7}{*}{PA}    
& H-GAIL  & 1.103$\pm$0.336& 1.218$\pm$0.339& 2.84\%& 0.116$\pm$0.015& 0.146$\pm$0.017& 0.47\%&  0.491&  33.842&  \textbf{0.006}&  0.431\\
& DRAIL   & 0.659$\pm$0.384& 0.697$\pm$0.371& 1.95\%& 0.118$\pm$0.009& 0.158$\pm$0.013& 0.47\%&  1.114&  32.756&  0.617&  0.655\\
& GAIL    & \underline{0.354$\pm$0.152}& \underline{0.406$\pm$0.138}& \underline{1.04\%}& 0.089$\pm$0.005& 0.127$\pm$0.013& 0.36\%&  0.844&  33.152&  0.347&  \underline{0.259}\\

& Transformer    & 1.259$\pm$0.048& 1.411$\pm$0.053& 3.73\%& \textbf{0.053$\pm$0.001}& \textbf{0.071$\pm$0.002}& \textbf{0.21\%}&  0.050&  30.993&  0.447&  2.418\\
& LSTM    & 1.125$\pm$0.467 & 1.247$\pm$0.494& 3.33\%& 0.086$\pm$0.031& 0.125$\pm$0.037& 0.30\%&  0.215&  32.527& 0.282  & 0.884 \\
& Diffusion    & 1.110$\pm$0.332 & 1.263$\pm$0.334 & 3.29\%& 0.098$\pm$0.008& 0.123$\pm$0.009& 0.40\%&  0.480&  31.651&  \underline{0.017}& 1.760 \\

& CRIL & \textbf{0.280$\pm$0.056}& \textbf{0.349$\pm$0.057}& \textbf{0.83\%}& \underline{0.071$\pm$0.007}& \underline{0.102$\pm$0.009}& \underline{0.28\%}&  0.544&  33.156&  0.047&  \textbf{0.255}\\ 
\hline

\multirow{7}{*}{MA}    
& H-GAIL  & 0.418$\pm$0.123& 0.527$\pm$0.125& 1.24\%& 0.093$\pm$0.009& 0.118$\pm$0.012& 0.38\%&  0.668&  33.677&  0.171&  0.265\\
& DRAIL   & 0.379$\pm$0.193& 0.417$\pm$0.179& 1.12\%& 0.121$\pm$0.011& 0.167$\pm$0.020& 0.49\%&  0.731&  33.747&  0.234&  0.336\\
& GAIL    & \underline{0.271$\pm$0.112}& \underline{0.334$\pm$0.108}& \underline{0.80\%}& 0.109$\pm$0.006& 0.155$\pm$0.009& 0.44\%&  0.702&  33.612&  0.205&  \underline{0.201}\\

& Transformer    & 1.200$\pm$0.063& 1.246$\pm$0.069& 3.55\%& \underline{0.085$\pm$0.001}& 0.139$\pm$0.001& \underline{0.32\%}&  0.043&  31.147&  0.454&  2.264\\
& LSTM& 0.774$\pm$0.362 & 0.863$\pm$0.367& 2.29\%& 0.112$\pm$0.021& 0.135$\pm$0.029& 0.35\%&  0.244&  32.388&  0.253 & 1.023 \\
& Diffusion    & 0.684$\pm$0.201 & 0.814$\pm$0.194 & 2.02\%& 0.092$\pm$0.010& \underline{0.114$\pm$0.013}& 0.37\%&  0.327&  32.334& \underline{0.170}& 1.077 \\

& CRIL & \textbf{0.251$\pm$0.064}& \textbf{0.315$\pm$0.066}& \textbf{0.74\%}& \textbf{0.075$\pm$0.005}& \textbf{0.111$\pm$0.007}& \textbf{0.30\%}&  0.643&  33.222&  \textbf{0.146}&  \textbf{0.189}\\ 
\hline

\multirow{7}{*}{BA}    
& H-GAIL  & \underline{0.772$\pm$0.361}& \underline{0.811$\pm$0.355}& \underline{2.29\%}& 0.108$\pm$0.021& 0.148$\pm$0.020& 0.44\%&  0.838&  34.016&  \underline{0.341}&  \underline{0.605}\\
& DRAIL   & 2.061$\pm$1.081& 2.093$\pm$1.080& 6.10\%& 0.105$\pm$0.029& 0.147$\pm$0.054& 0.42\%&  1.082&  32.718&  0.585&  0.693\\
& GAIL    & 0.913$\pm$0.229& 0.937$\pm$0.226& 2.70\%& \underline{0.099$\pm$0.009}& \underline{0.140$\pm$0.013}& \underline{0.40\%}&  1.036&  32.593&  0.539&  0.818\\

& Transformer    & 7.366$\pm$0.272& 8.365$\pm$0.311& 21.83\%& 0.413$\pm$0.017& 0.418$\pm$0.016& 1.67\%&  0.056&  19.075&  0.441&  14.336\\
& LSTM& 10.815$\pm$11.323 & 16.420$\pm$20.664& 32.13\%& 1.361$\pm$2.028& 2.079$\pm$3.069& 5.50\%&  10.587&  31.420& 10.090  & 1.991 \\
& Diffusion    & 3.201$\pm$0.821 & 3.545$\pm$1.030 & 9.47\%& 0.472$\pm$0.112& 0.591$\pm$0.140& 1.90\%&  1.905&  28.933& 1.408 & 4.478 \\

& CRIL & \textbf{0.214$\pm$0.039}& \textbf{0.262$\pm$0.041}& \textbf{0.63\%}& \textbf{0.093$\pm$0.003}& \textbf{0.131$\pm$0.005}& \textbf{0.37\%}&  0.782&  33.641&  \textbf{0.285}&  \textbf{0.230}\\ 
\hline

\multirow{7}{*}{MaA}  
& H-GAIL  & 0.376$\pm$0.254& 0.480$\pm$0.277& 1.11\%& 0.098$\pm$0.039& \underline{0.113$\pm$0.043}& \underline{0.40\%}&  0.323&  32.988&  0.174&  0.423\\
& DRAIL   & \underline{0.291$\pm$0.273}& \textbf{0.322$\pm$0.255}& \underline{0.86\%}& 0.106$\pm$0.017& 0.119$\pm$0.023& 0.43\%&  0.664&  33.563&  \underline{0.167}&  \underline{0.152}\\
& GAIL    & 0.346$\pm$0.325& 0.458$\pm$0.331& 1.02\%&\underline{ 0.104$\pm$0.011}& 0.118$\pm$0.016& 0.42\%&  0.675&  33.219&  0.178&  0.192\\

& Transformer    & 4.548$\pm$0.279& 5.001$\pm$0.307& 13.49\%& 0.242$\pm$0.015& 0.260$\pm$0.014& 0.97\%&  0.053&  34.610&  0.444&  1.199\\
& LSTM& 1.357$\pm$0.302 & 1.511$\pm$0.320& 4.02\%& 0.124$\pm$0.012& 0.139$\pm$0.013& 0.50\%&  0.024&  34.009&  0.473 & 0.598 \\
& Diffusion    & 0.784$\pm$0.139 & 0.805$\pm$0.147 & 2.22\%& 0.112$\pm$0.010& 0.140$\pm$0.010& 0.45\%&  0.281&  33.872& 0.216 & 0.461 \\

& CRIL & \textbf{0.287$\pm$0.219}& \underline{0.335$\pm$0.237}& \textbf{0.85\%}& \textbf{0.096$\pm$0.010}& \textbf{0.109$\pm$0.009}& \textbf{0.39\%}&  0.336&  33.329&  \textbf{0.161}&  \textbf{0.082}\\ 
\hline

\multirow{7}{*}{DPA}  
& H-GAIL  & 0.750$\pm$0.542& 0.807$\pm$0.574& 2.23\%& \underline{0.100$\pm$0.028}& 0.122$\pm$0.035& \underline{0.40\%}&  0.315&  33.930&  0.182&  0.519\\
& DRAIL   & \underline{0.410$\pm$0.292}& \underline{0.523$\pm$0.276}& \underline{1.25\%}& 0.103$\pm$0.002& \underline{0.115$\pm$0.003}& 0.42\%&  0.598&  33.022&  \underline{0.101}&  0.389\\
& GAIL    & 0.480$\pm$0.285& 0.568$\pm$0.277& 1.41\%& \underline{0.100$\pm$0.003}& \textbf{0.112$\pm$}0.002& \underline{0.40\%}&  0.357&  33.078&  0.140&  0.333\\

& Transformer    & 4.128$\pm$0.080& 4.671$\pm$0.090& 12.24\%& 0.297$\pm$0.005& 0.303$\pm$0.005& 1.19\%&  0.042&  34.107&  0.455&  0.696\\
& LSTM& 12.329$\pm$17.523 & 17.974$\pm$27.486& 36.61\%& 1.489$\pm$2.347& 1.940$\pm$3.233& 6.01\%&  8.070&  34.057& 7.573  & 0.646 \\
& Diffusion    & 1.296$\pm$0.368& 1.429$\pm$0.365 & 3.85\%& 0.145$\pm$0.009& 0.192$\pm$0.017& 0.59\%&  0.628&  33.905&0.131  &\underline{0.316}  \\

& CRIL & \textbf{0.408$\pm$0.254}& \textbf{0.465$\pm$0.246}& \textbf{1.21\%}& \textbf{0.097$\pm$0.001}& \textbf{0.112$\pm$0.001}& \textbf{0.39\%}&  0.399&  33.706&  \textbf{0.098}&  \textbf{0.295}\\ 
\hline

\end{tabular}
}
\end{table*}

\subsection{Performance evaluation under cyberattack scenarios}\label{results analysis}

As shown in Table~\ref{Numerical comparison of different methods}, CRIL achieves the best overall recovery performance across the evaluated cyberattack scenarios. Although several baselines obtain lower velocity errors in specific cases, such as Transformer under FA and PA, their spacing recovery is substantially less accurate, indicating that velocity reconstruction alone is insufficient to restore the complete longitudinal car-following state. CRIL consistently achieves the lowest or near-lowest spacing errors while maintaining competitive velocity accuracy. 

Regarding post-recovery stability and safety, CRIL also yields consistently low $\Delta R_v$ and $\Delta s_{\min}$, where smaller values indicate closer agreement with the attack-free reference in speed-fluctuation propagation and minimum spacing, respectively. Although individual baselines perform better on specific metrics or under particular attacks, CRIL provides the most balanced overall performance in trajectory accuracy, traffic stability, and operational safety. This advantage is attributed to its cross-branch learning strategy, which jointly captures nominal driving behavior and enhances robustness to attack-induced distribution shifts.

\subsection{State-Space Distribution Analysis}

\begin{figure*}[h]
\centering

\subfloat[BA ]{
\includegraphics[width=0.319\textwidth]{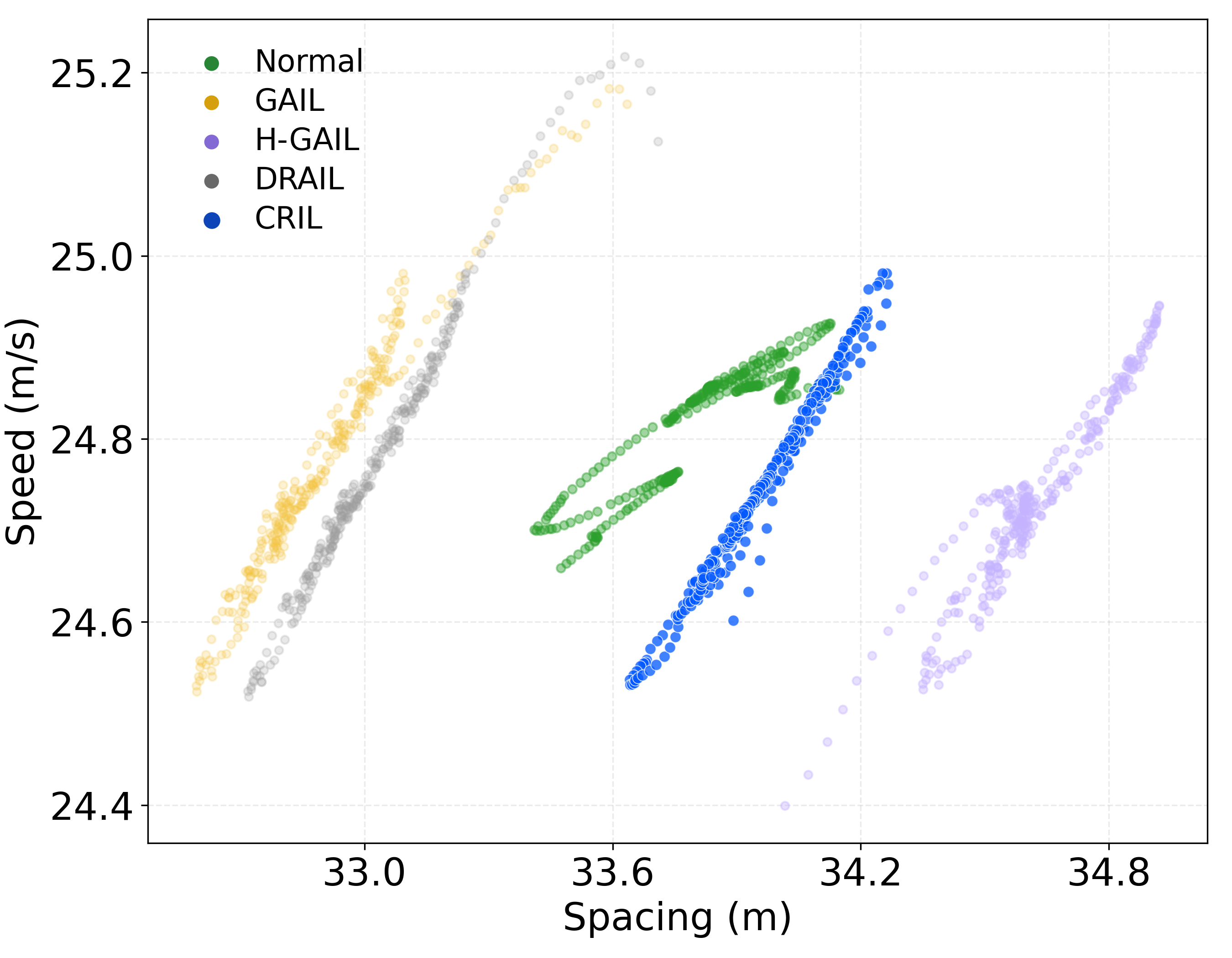}
}
\hfill
\subfloat[FA ]{
\includegraphics[width=0.319\textwidth]{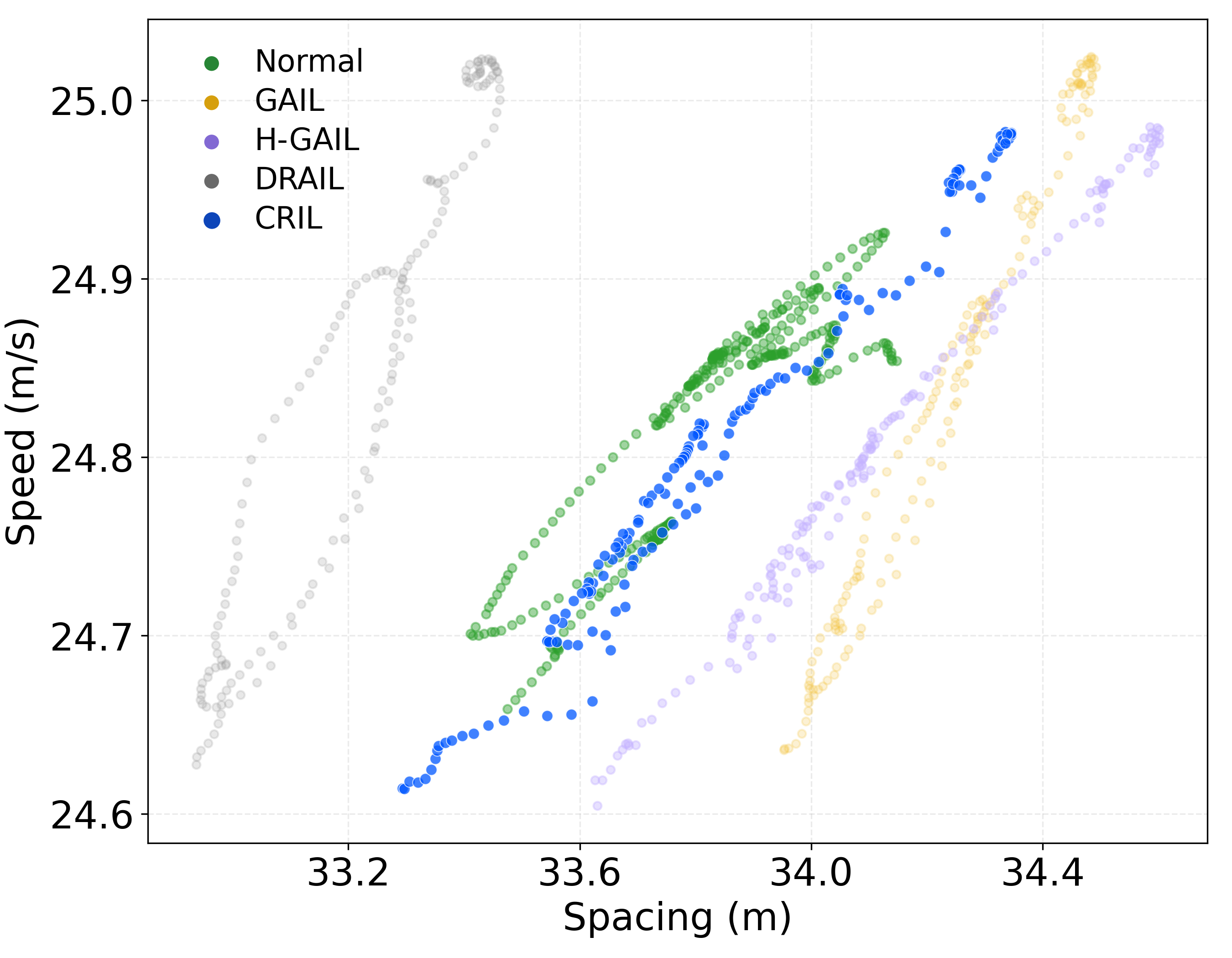}
}
\hfill
\subfloat[MA ]{
\includegraphics[width=0.319\textwidth]{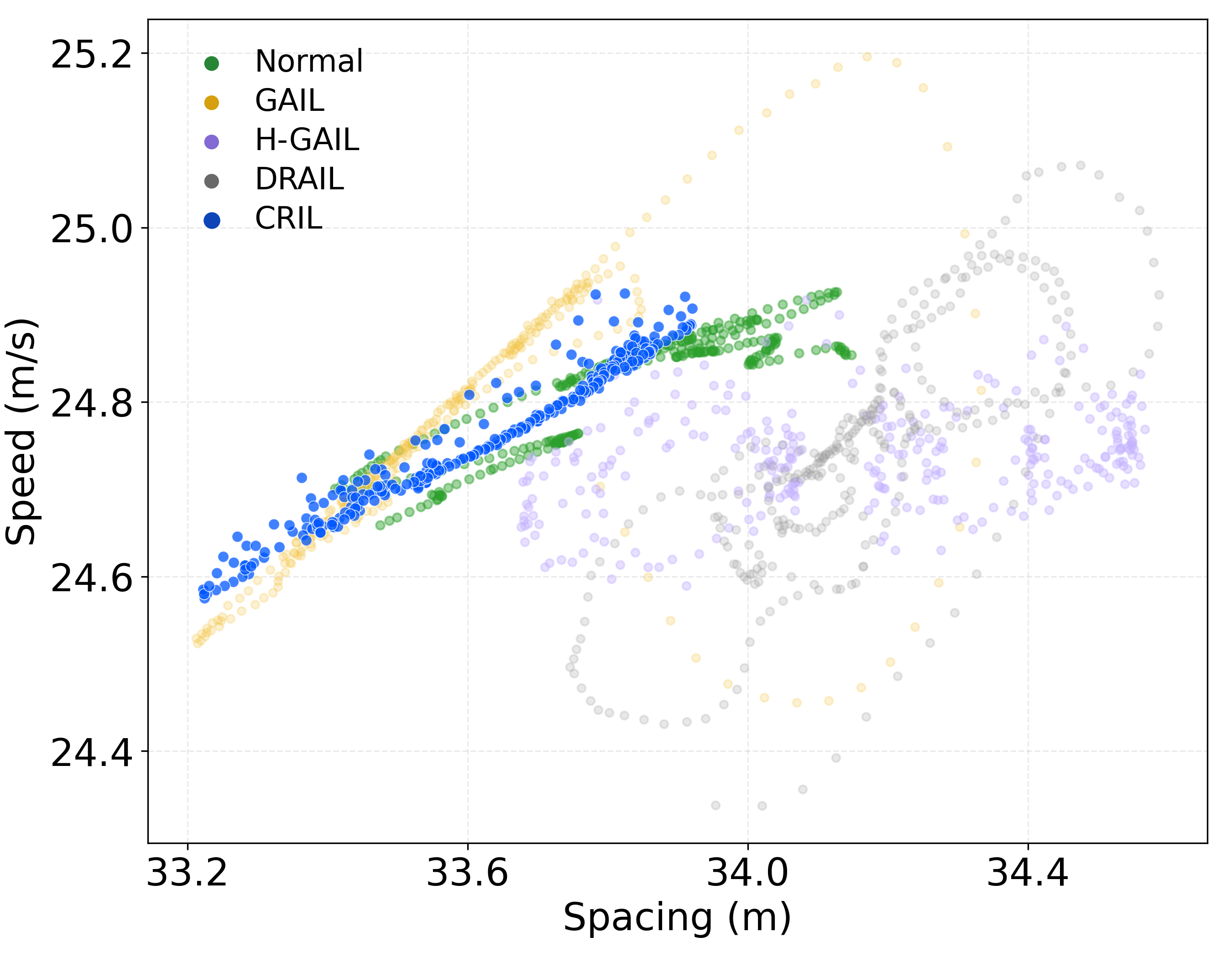}
}

\vspace{0.15cm}

\subfloat[PA ]{
\includegraphics[width=0.319\textwidth]{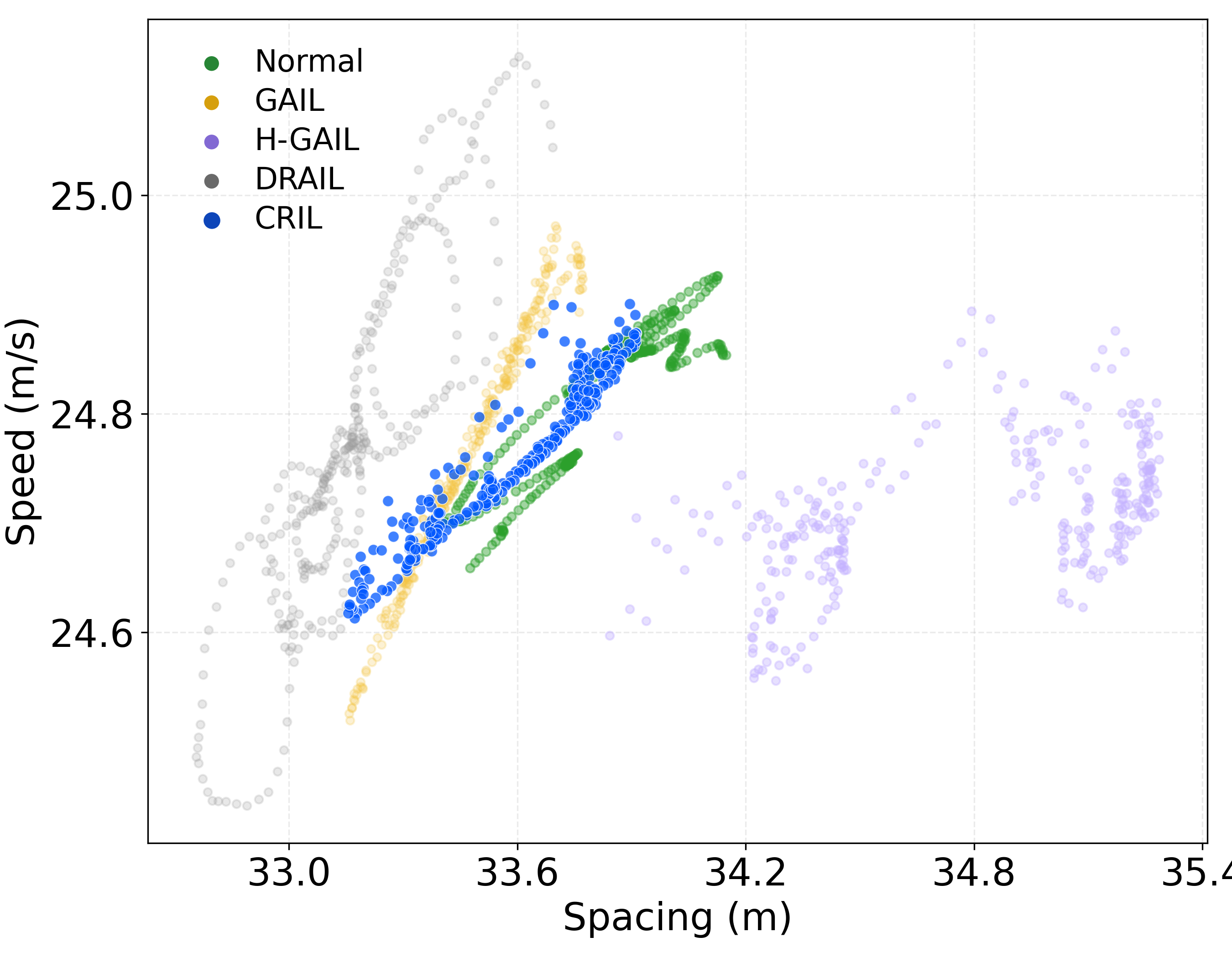}
}
\hfill
\subfloat[MaA]{
\includegraphics[width=0.319\textwidth]{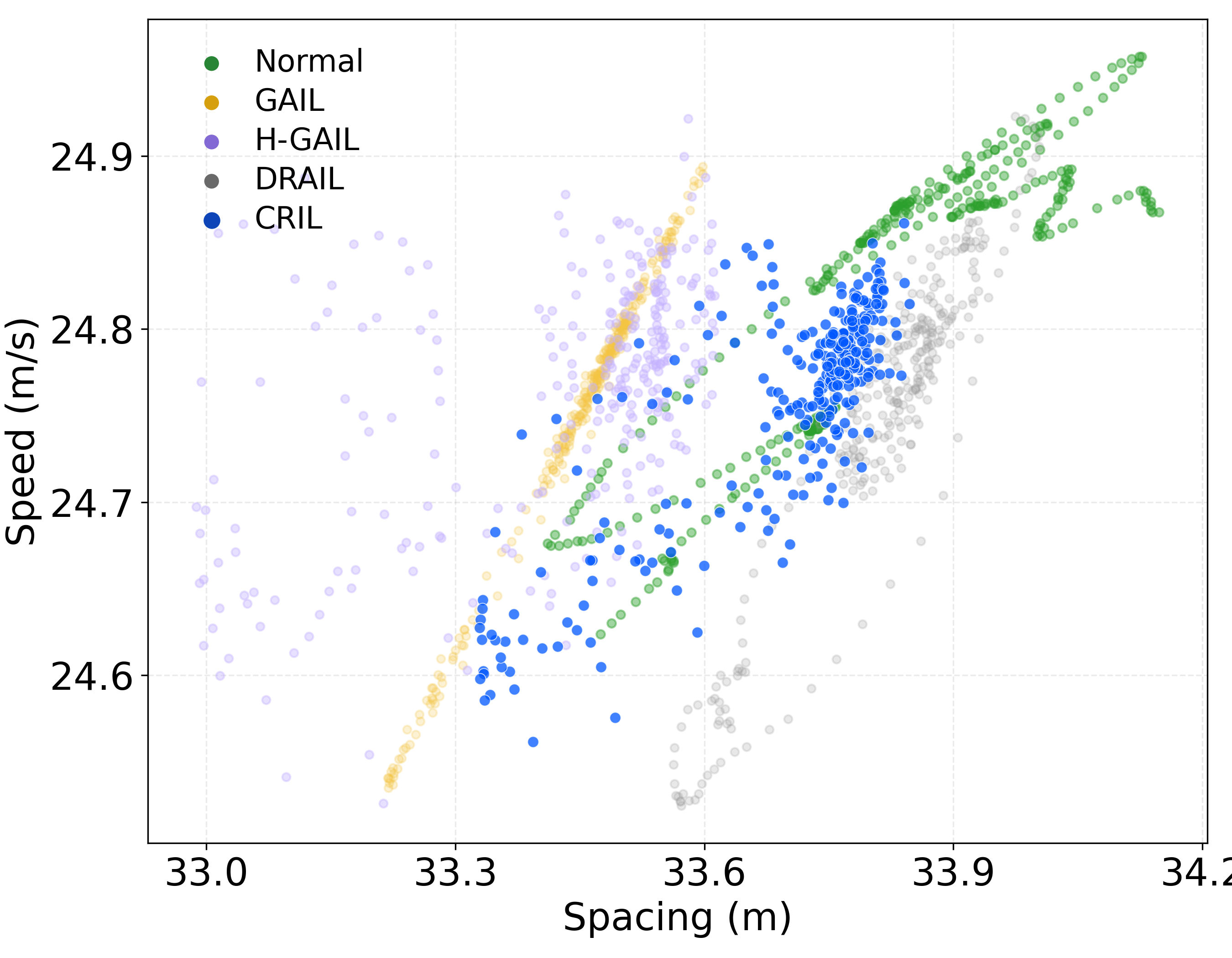}
}
\hfill
\subfloat[DPA]{
\includegraphics[width=0.319\textwidth]{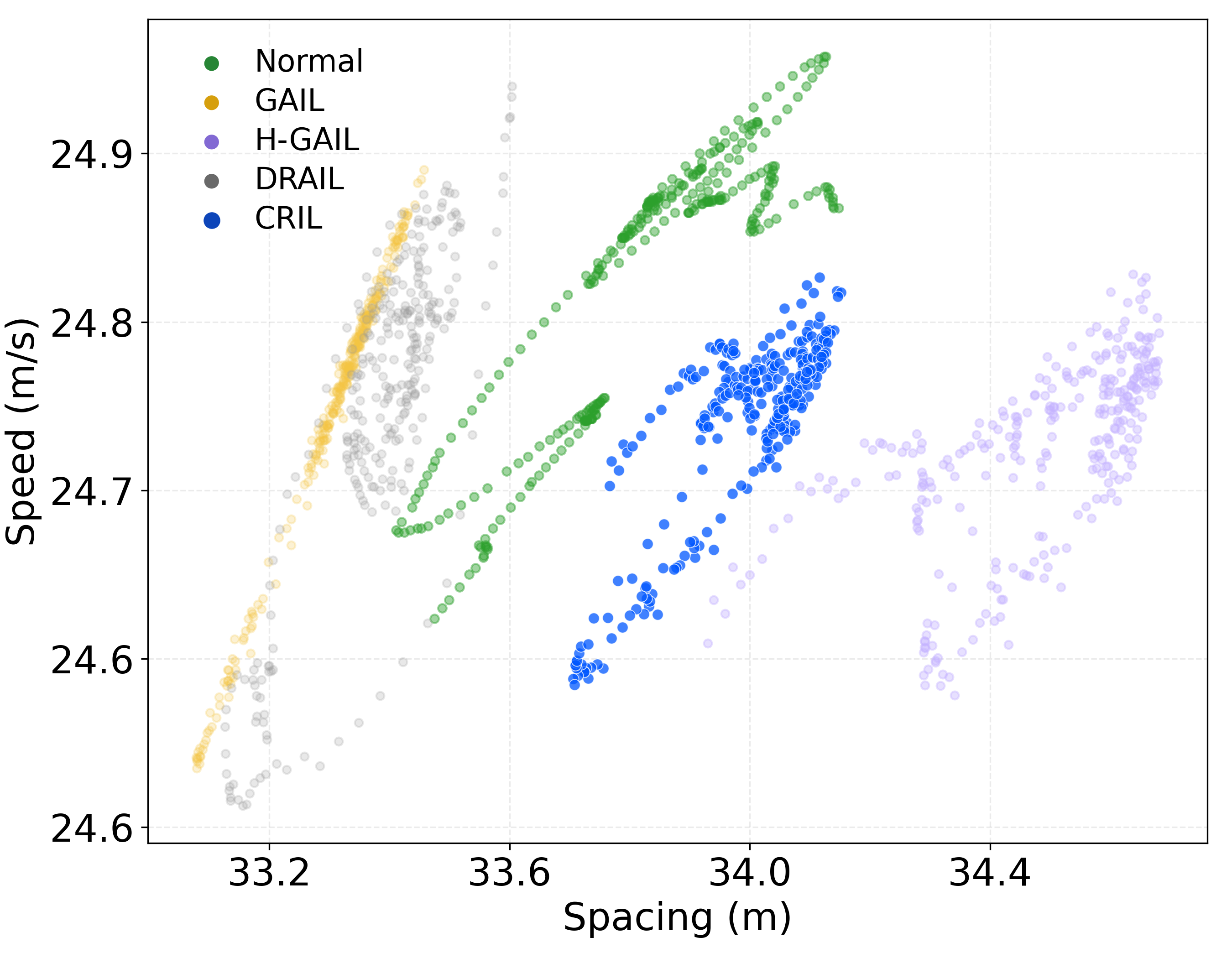}
}

\caption{
Comparison of state-space distributions between attack-free and recovered trajectories under six cyberattack scenarios.
}
\label{fig:state_distribution_comparison}
\end{figure*}

To assess whether the recovered trajectories return to the nominal driving-state domain, Fig. \ref{fig:state_distribution_comparison} compares the state-space distributions produced by different methods under six cyberattack scenarios.
The baseline methods exhibit substantial distribution shifts, including displaced, over-dispersed, and locally distorted state clusters, indicating incomplete restoration of the spacing–speed relationship. 

CRIL shows substantial overlap with the attack-free distribution under FA, MA, and PA, and markedly reduces the distribution gap under BA and DPA. Although a residual shift remains under MaA, CRIL generally preserves the shape and dependence structure of the nominal state distribution more effectively than the compared methods. These results confirm that the cross-branch learning strategy mitigates attack-induced distribution shifts and improves the restoration of rational car-following dynamics, rather than merely reducing pointwise trajectory errors.

\subsection{Ablation study}

\begin{table}[htbp]
\centering
\caption{Ablation results for CRIL under different cyberattack scenarios}
\label{ablation results}

\scriptsize
\renewcommand{\arraystretch}{1.3}
\setlength{\tabcolsep}{1.5pt}

\begin{tabular}{>{\centering\arraybackslash}m{0.6cm}
                >{\centering\arraybackslash}m{1.6cm}|
                cccccc}

\hline

\multicolumn{2}{c|}{}
& \multicolumn{3}{c}{Spacing}
& \multicolumn{3}{c}{Speed} \\

\cline{3-8}

\multirow{2}{*}[3ex]{Attack}
& \multirow{2}{*}[3ex]{Method}
& $\text{MAE}_s$
& $\text{RMSE}_s$
& $\text{MAPE}_s$
& $\text{MAE}_v$
& $\text{RMSE}_v$
& $\text{MAPE}_v$ \\

\hline

\multirow{4}{*}{FA}
& CRIL& \textbf{0.317}& \textbf{0.391}& \textbf{0.94\%}& \textbf{0.121}& \textbf{0.148}& \textbf{0.51\%}\\
& clean branch&0.452&0.528&1.34\%&0.137&0.169&0.56\%\\
& noise branch&0.393&0.470&1.16\%&0.133&0.162&0.54\%\\
& single branch&0.466&0.547&1.38\%&0.130&0.155&0.52\%\\

\hline

\multirow{4}{*}{PA}
& CRIL& \textbf{0.280}& \textbf{0.349}& \textbf{0.83\%}& \textbf{0.071}& \textbf{0.102}& \textbf{0.28\%}\\
& clean branch&1.103&1.218&2.84\%&0.116&0.146&0.47\%\\
& noise branch&0.386&0.450&1.14\%&0.073&0.105&0.30\%\\
& single branch&0.394&0.474&1.17\%&0.090&0.119&0.36\%\\

\hline

\multirow{4}{*}{MA}
& CRIL& \textbf{0.251}& \textbf{0.315}& \textbf{0.74\%}& \textbf{0.075}& \textbf{0.111}& \textbf{0.30\%}\\
& clean branch&0.650&0.723&1.93\%&0.089&0.115&0.36\%\\
& noise branch&1.095&1.135&3.24\%&0.117&0.163&0.47\%\\
& single branch&1.322&1.365&3.91\%&0.081&0.114&0.35\%\\

\hline

\multirow{4}{*}{BA}
& CRIL& \textbf{0.214}& \textbf{0.262}& \textbf{0.63\%}& \textbf{0.093}& \textbf{0.131}& \textbf{0.37\%}\\
& clean branch&0.613&0.653&1.82\%&0.104&0.149&0.42\%\\
& noise branch&0.512&0.832&1.51\%&0.146&0.298&0.59\%\\
& single branch&0.973&0.997&2.88\%&0.112&0.159&0.45\%\\

\hline

\multirow{4}{*}{MaA}
& CRIL& \textbf{0.287}& \textbf{0.335}& \textbf{0.85\%}& \textbf{0.096}& \textbf{0.109}& \textbf{0.39\%}\\
& clean branch&1.094&1.155&3.23\%&0.107&0.120&0.43\%\\
& noise branch&0.314&0.354&0.93\%&\textbf{0.096}&0.111&\textbf{0.39\%}\\
& single branch&1.141&1.200&3.37\%&0.103&0.116&0.42\%\\

\hline

\multirow{4}{*}{DPA}
& CRIL& \textbf{0.408}& \textbf{0.465}& \textbf{1.21\%}& \textbf{0.097}& \textbf{0.112}& \textbf{0.39\%}\\
& clean branch&0.700&0.764&2.06\%&0.098&0.127&0.40\%\\
& noise branch&0.609&0.653&1.81\%&\textbf{0.097}&0.117&\textbf{0.39\%}\\
& single branch&0.432&0.534&1.27\%&0.109&0.124&0.44\%\\

\hline

\end{tabular}

\end{table}

To evaluate the contribution of the cross-branch design, four variants were compared: the full CRIL model, clean branch, noise branch, and single-branch model trained on both clean and perturbed samples. All results were averaged over five independent runs under the settings described in Section \ref{setup}.

As shown in Table \ref{ablation results}, the full CRIL model achieves the best overall performance across the six attack scenarios. Removing the noise-enhanced branch generally reduces recovery accuracy, demonstrating the benefit of perturbation-aware learning under corrupted observations. Conversely, the noise-enhanced-only variant remains inferior to the full model, indicating that robustness-oriented training alone is insufficient to preserve nominal driving behavior. The single-branch variant also performs less effectively, showing that jointly mixing clean and perturbed samples cannot fully exploit their complementary characteristics. These results validate the dedicated roles of the clean branch in learning nominal driving behavior and the noise-enhanced branch in improving recovery robustness.

\subsection{Parameter sensitivity analysis}

\begin{figure*}[!t]
\centering
\captionsetup[subfigure]{skip=1pt}

\begin{subfigure}{0.48\textwidth}
\centering
\includegraphics[width=\textwidth]{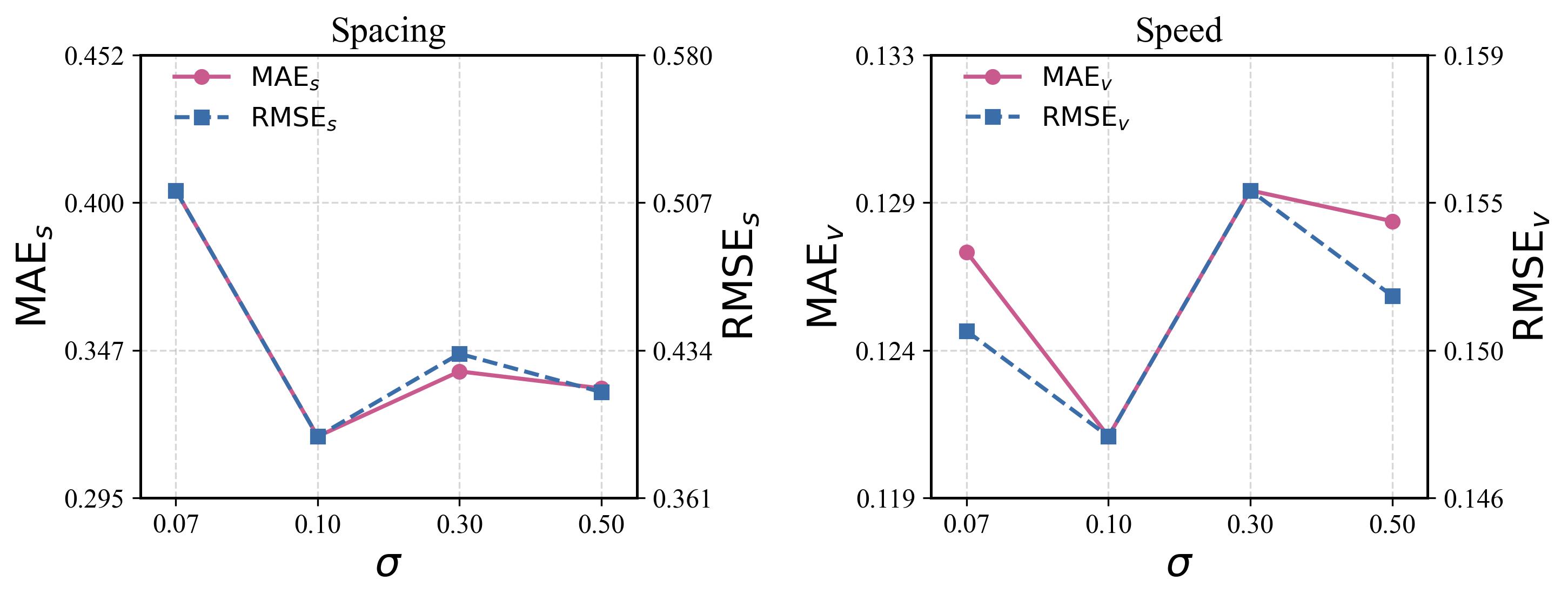}
\caption{FA}
\end{subfigure}
\hfill
\begin{subfigure}{0.48\textwidth}
\centering
\includegraphics[width=\textwidth]{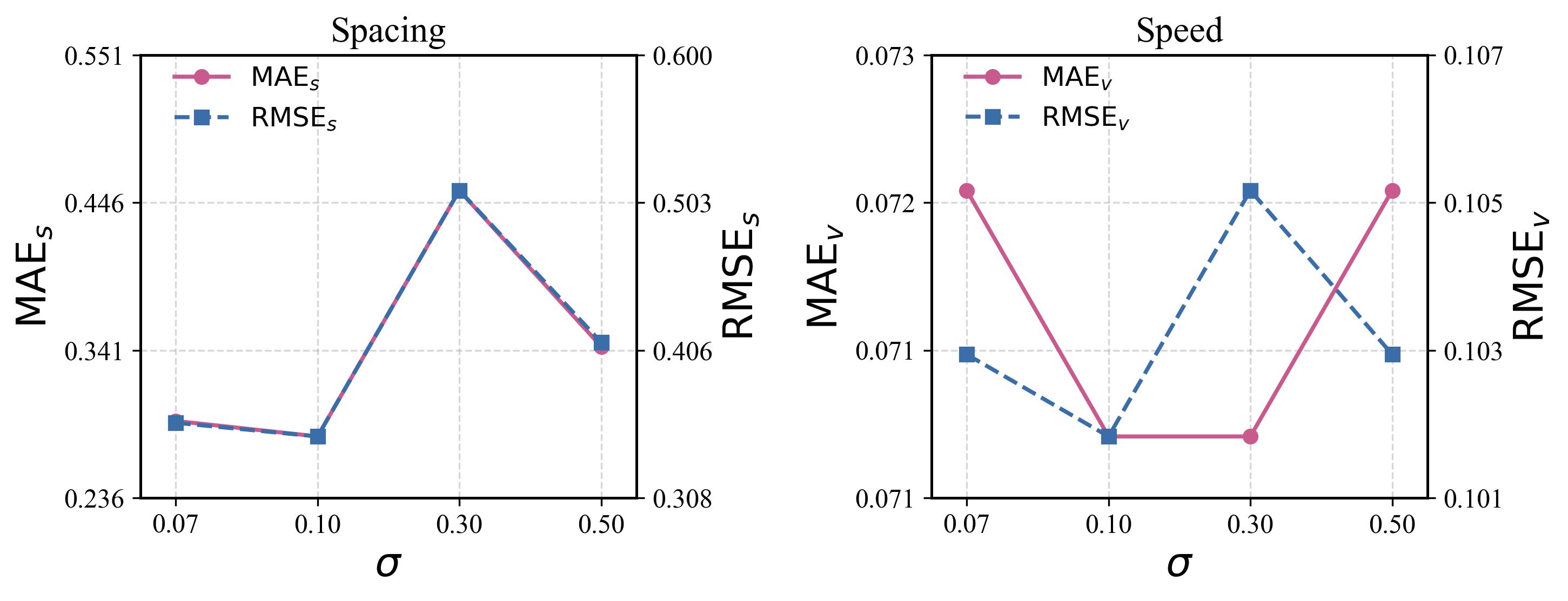}
\caption{PA}
\end{subfigure}

\vspace{0em}

\begin{subfigure}{0.48\textwidth}
\centering
\includegraphics[width=\textwidth]{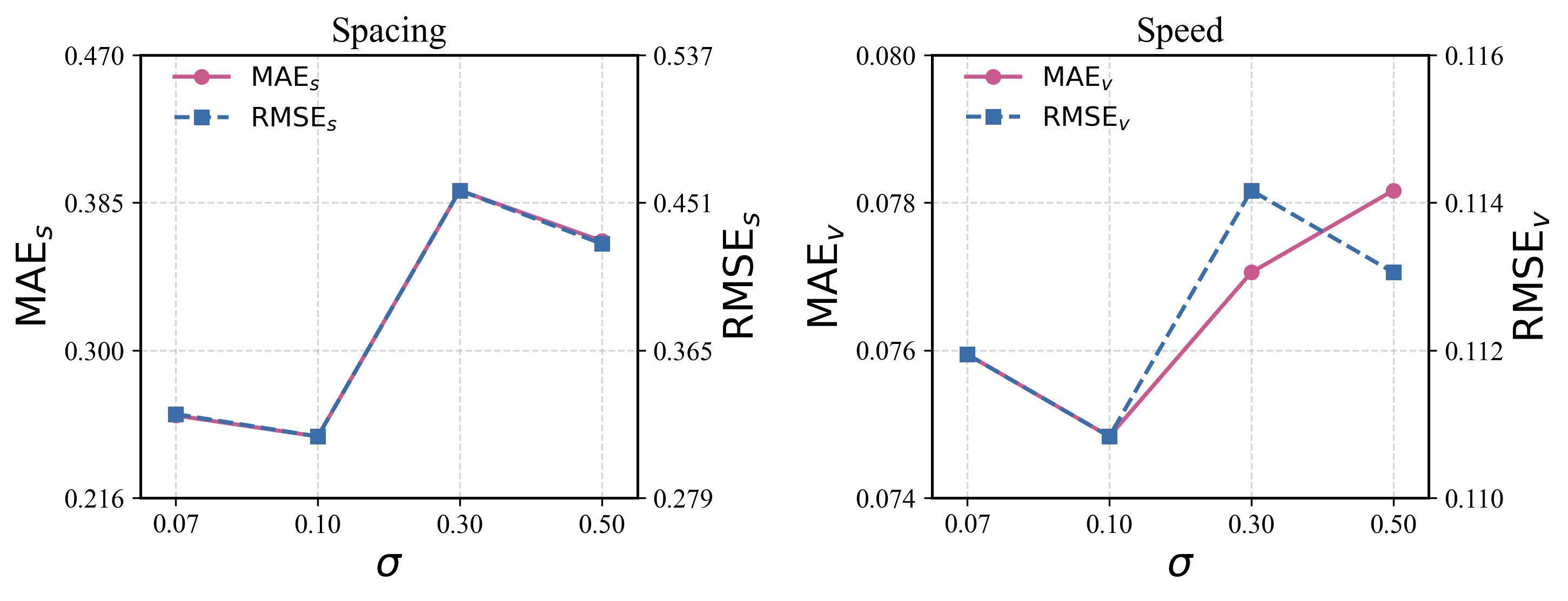}
\caption{MA}
\end{subfigure}
\hfill
\begin{subfigure}{0.48\textwidth}
\centering
\includegraphics[width=\textwidth]{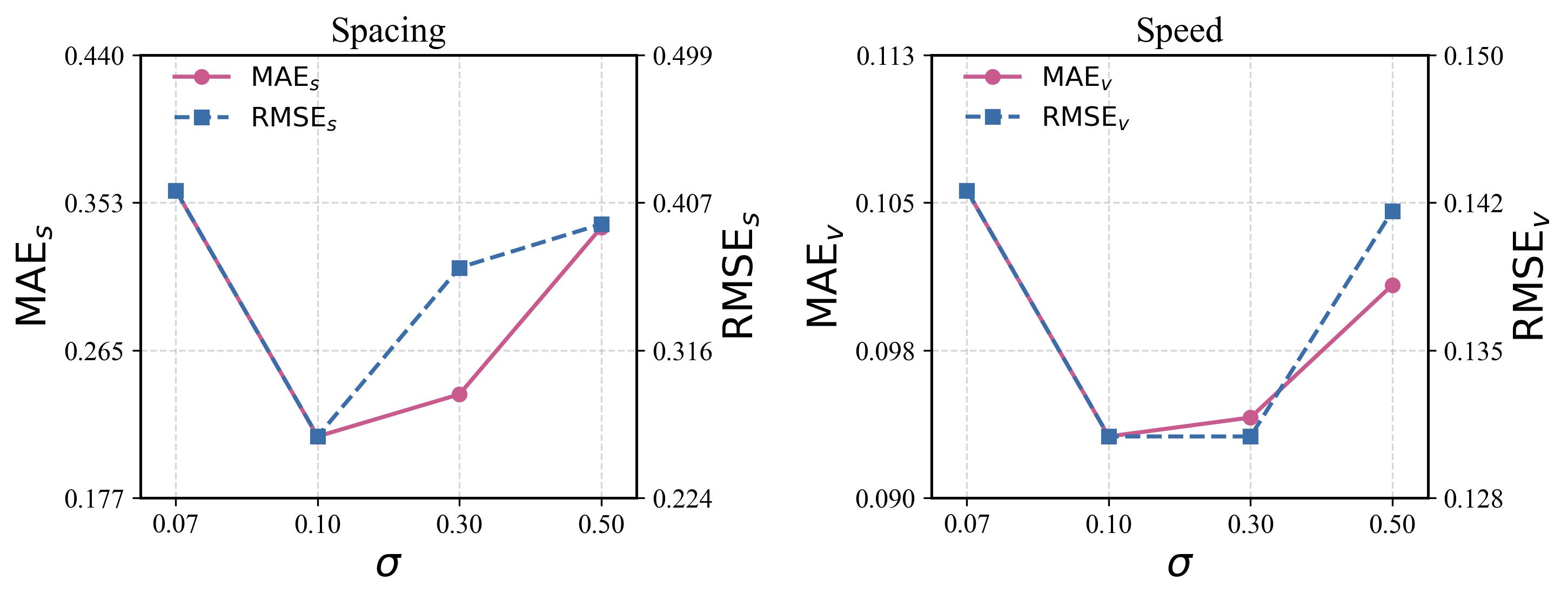}
\caption{BA}
\end{subfigure}

\vspace{0em}

\begin{subfigure}{0.48\textwidth}
\centering
\includegraphics[width=\textwidth]{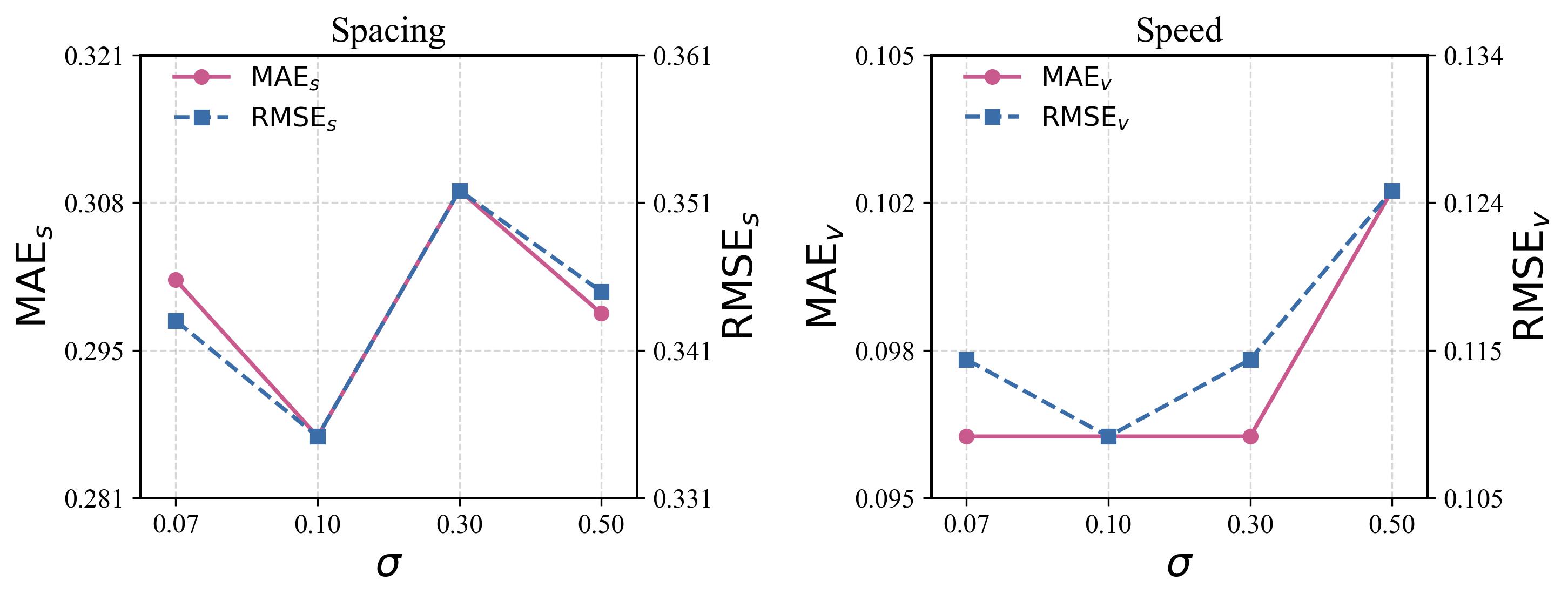}
\caption{MaA}
\end{subfigure}
\hfill
\begin{subfigure}{0.48\textwidth}
\centering
\includegraphics[width=\textwidth]{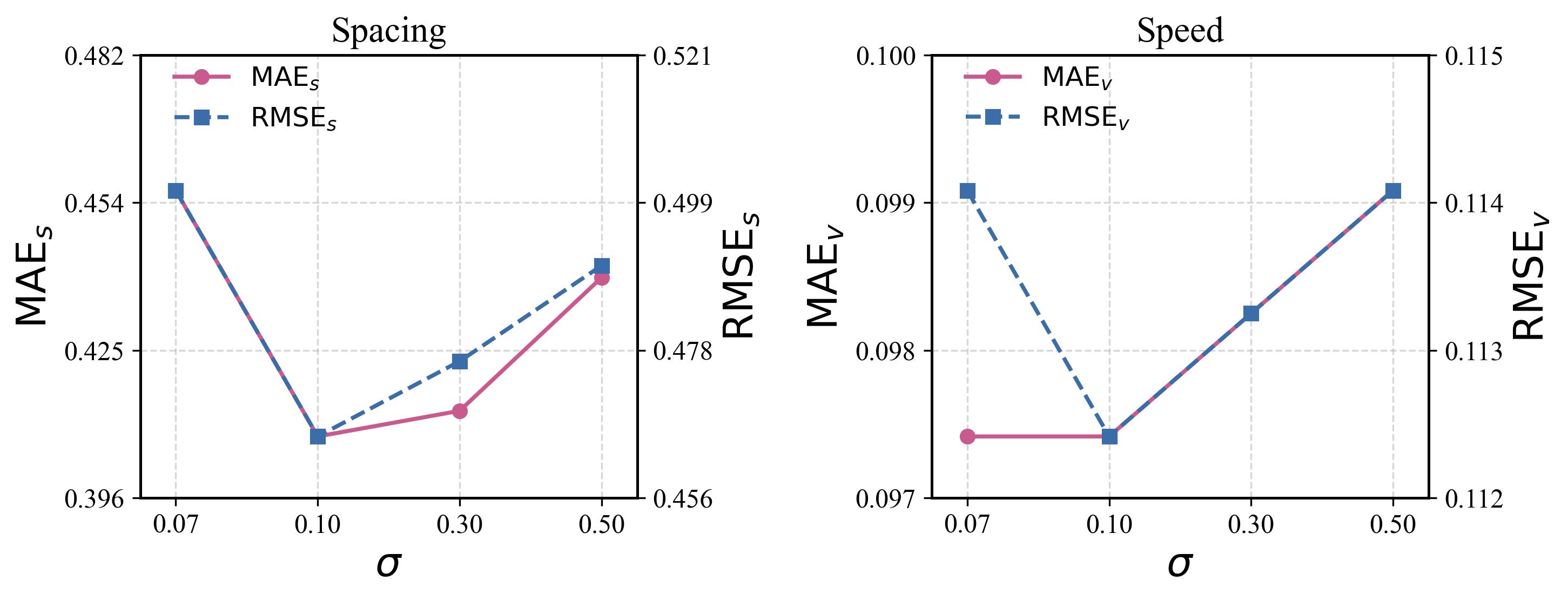}
\caption{DPA}
\end{subfigure}

\caption{Sensitivity analysis of CRIL to increasing Gaussian noise standard deviations under six cyberattack scenarios.}
\label{fig:noise_level_all}
\end{figure*}

\begin{figure*}[!t]
\centering

\begin{subfigure}{0.48\textwidth}
\centering
\includegraphics[width=\textwidth]{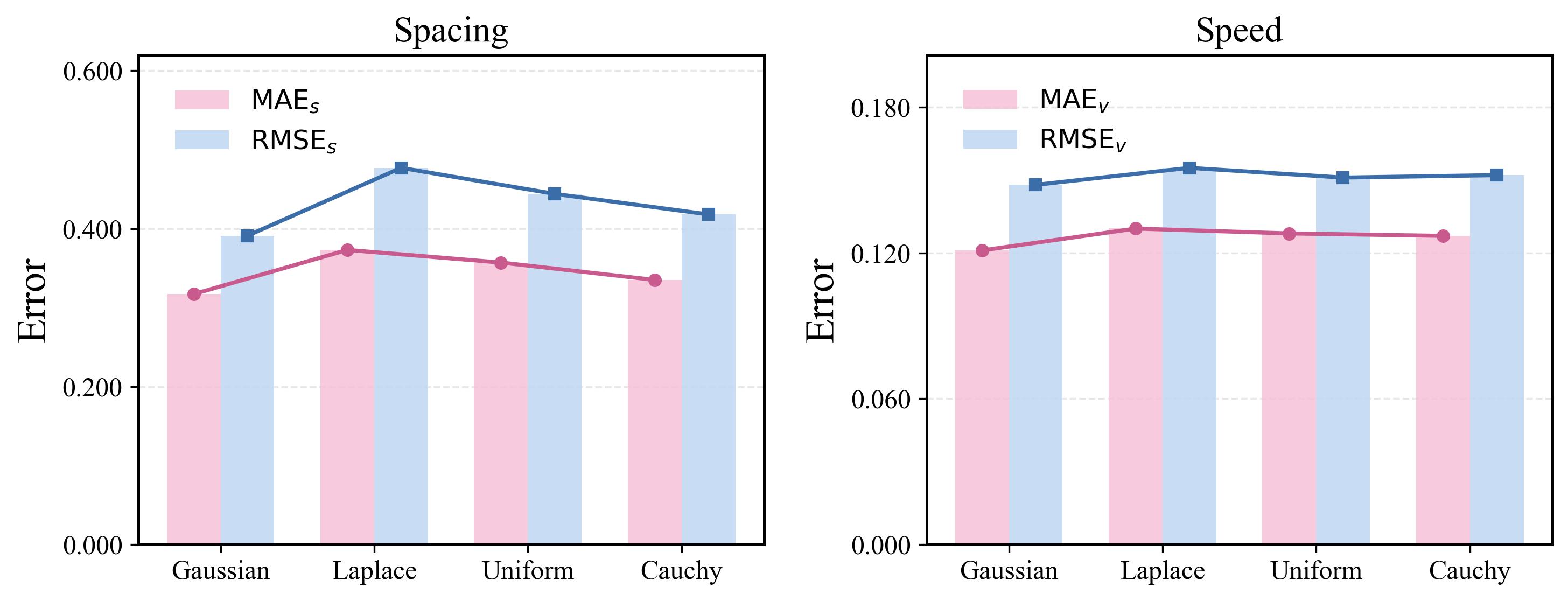}
\caption{FA}
\end{subfigure}
\hfill
\begin{subfigure}{0.48\textwidth}
\centering
\includegraphics[width=\textwidth]{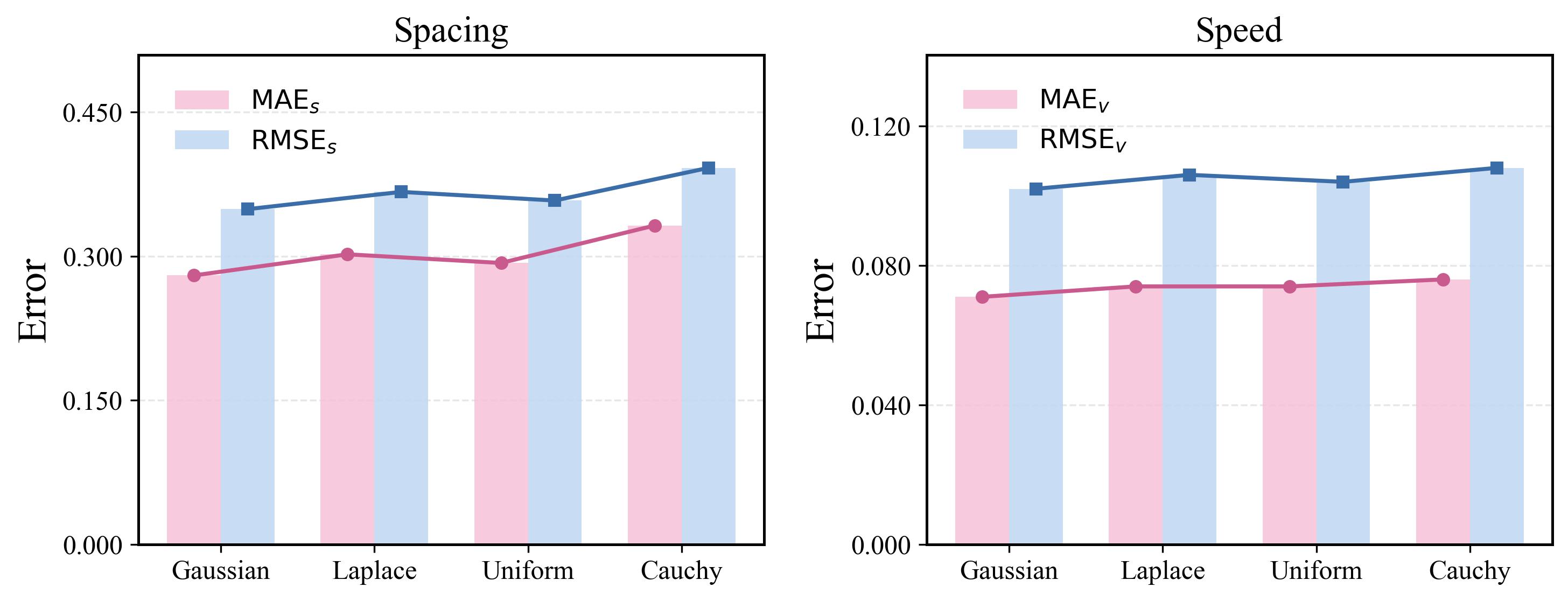}
\caption{PA}
\end{subfigure}

\vspace{-0.1em}
\begin{subfigure}{0.48\textwidth}
\centering
\includegraphics[width=\textwidth]{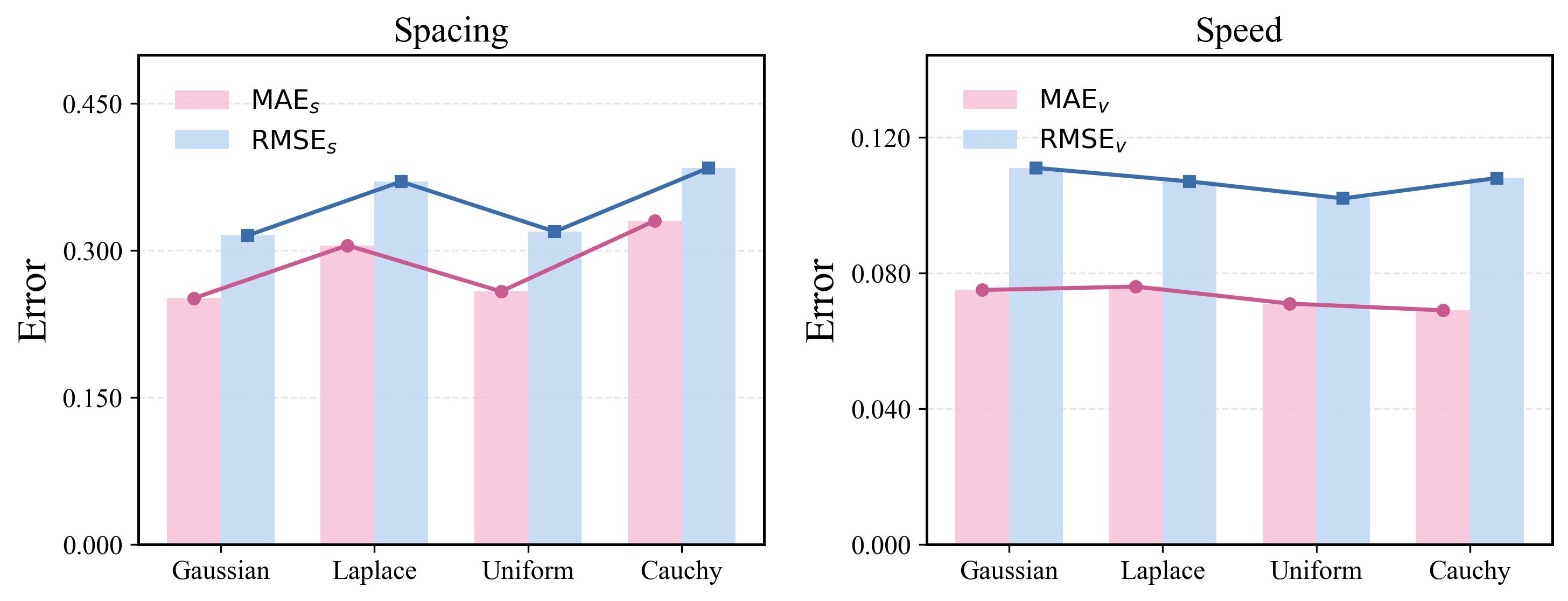}
\caption{MA}
\end{subfigure}
\hfill
\begin{subfigure}{0.48\textwidth}
\centering
\includegraphics[width=\textwidth]{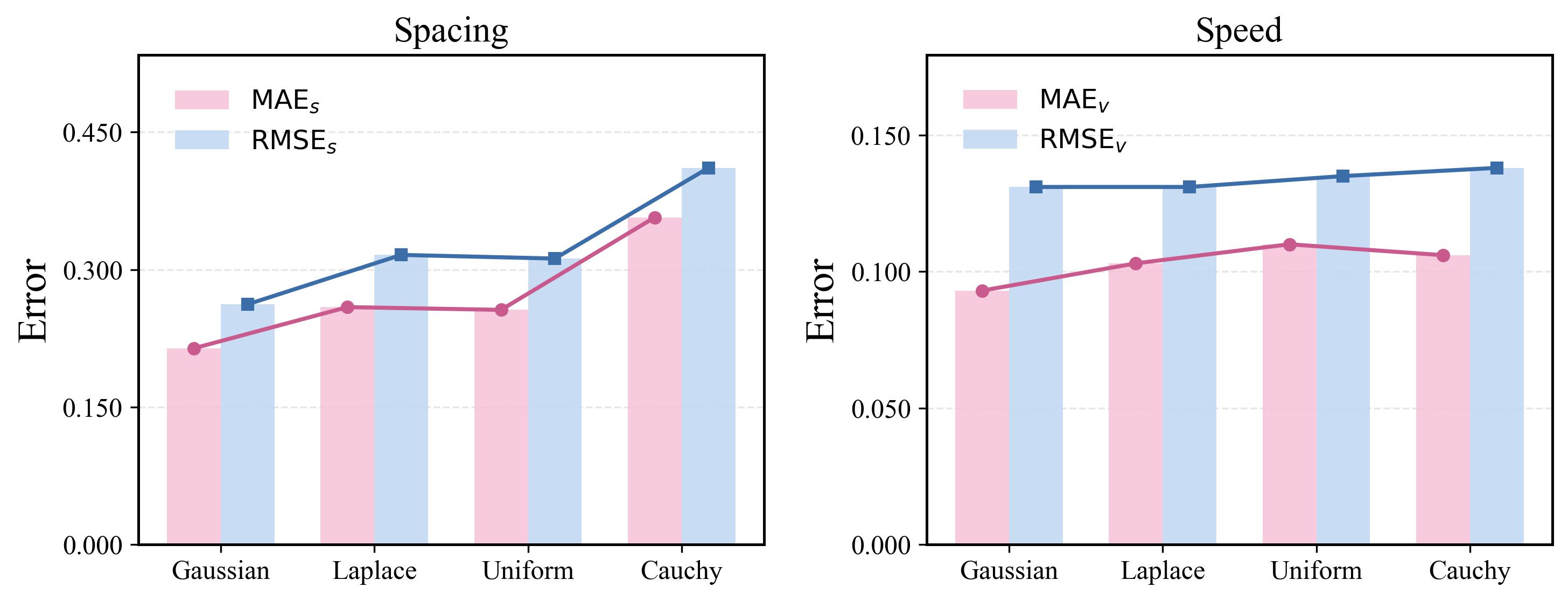}
\caption{BA}
\end{subfigure}

\vspace{-0.1em}
\begin{subfigure}{0.48\textwidth}
\centering
\includegraphics[width=\textwidth]{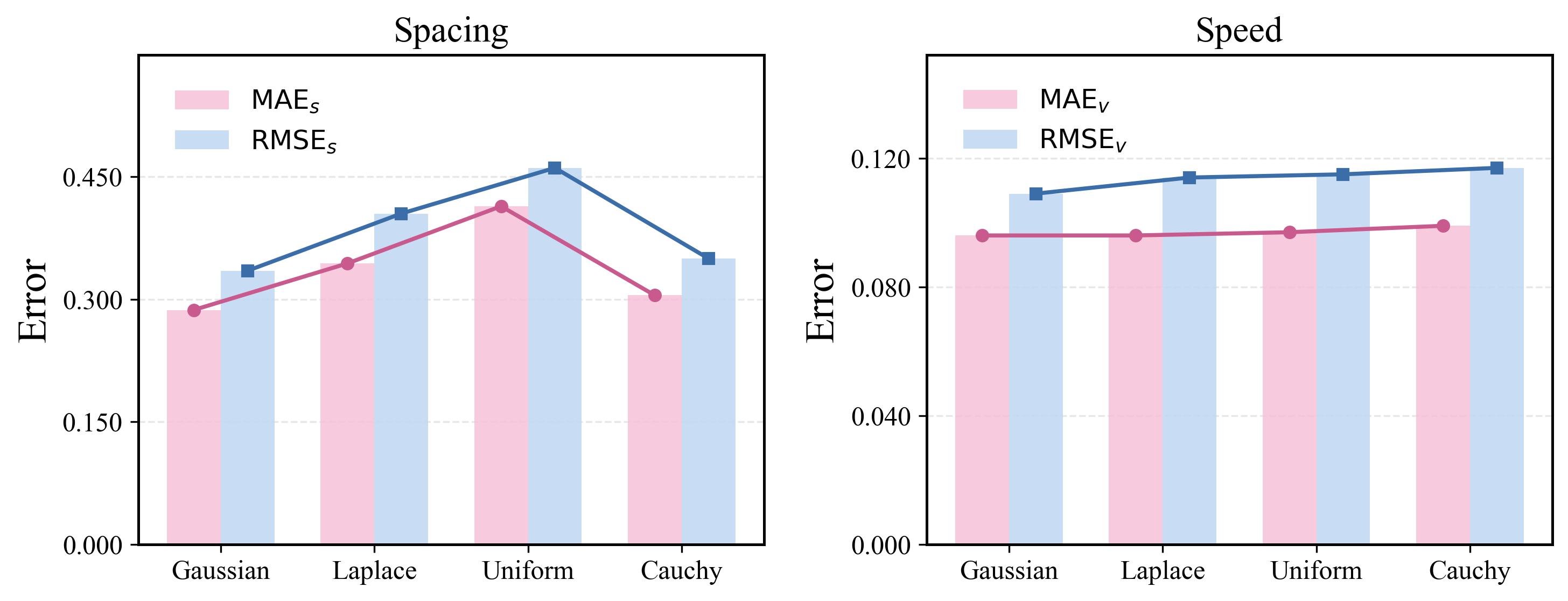}
\caption{MaA}
\end{subfigure}
\hfill
\begin{subfigure}{0.48\textwidth}
\centering
\includegraphics[width=\textwidth]{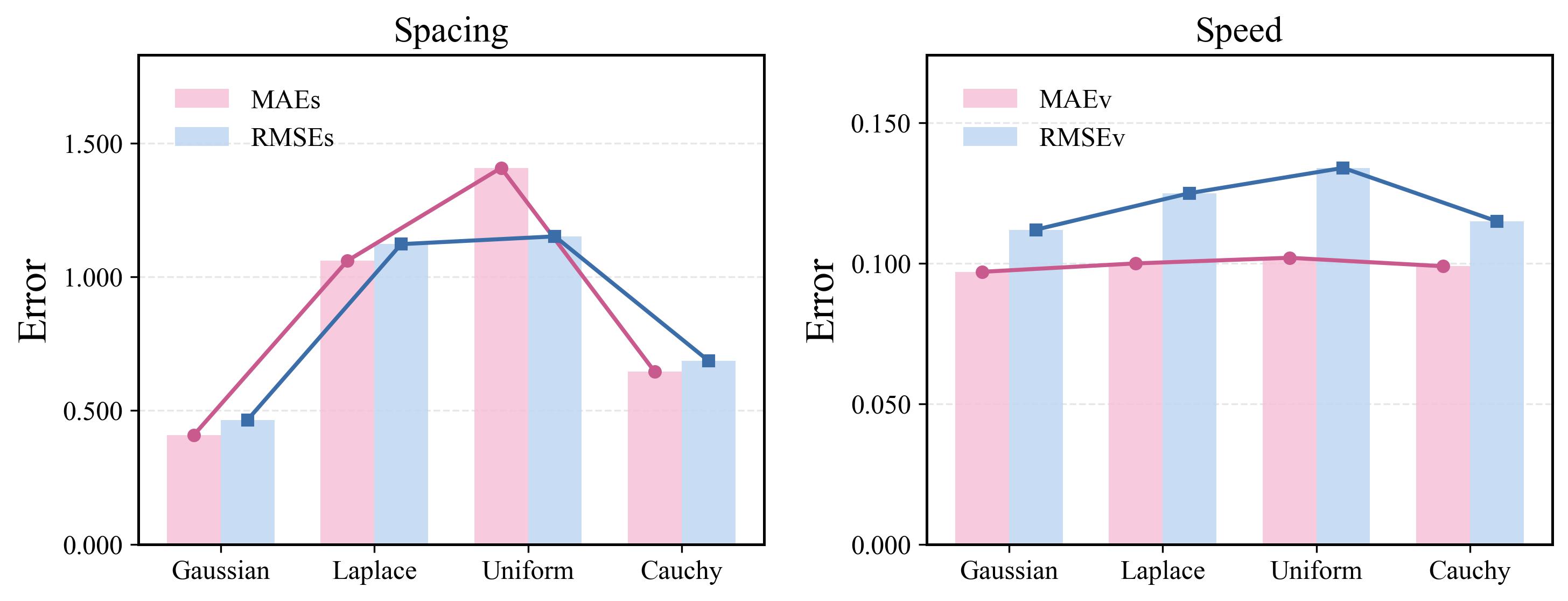}
\caption{DPA}
\end{subfigure}

\caption{
Sensitivity analysis of CRIL to different noise distributions at a fixed noise level of 0.1 under six cyberattack scenarios.
}
\label{fig:noise_different_type}

\end{figure*}

The sensitivity of CRIL to the perturbation configuration was evaluated in terms of noise intensity and distribution type. All experiments followed the settings in Section \ref{setup}, with only the perturbation configuration varied.
Fig. \ref{fig:noise_level_all} compares Gaussian perturbations with increasing standard deviations. Although the optimal intensity varies slightly across attacks and metrics, a standard deviation of $0.1$ achieves the most consistent speed and spacing recovery performance, particularly under FA, PA, BA, MA, and MaA. Larger perturbations generally increase recovery errors, indicating that excessive noise may distort the nominal driving information required for policy learning.

Fig. \ref{fig:noise_different_type} further compares Gaussian, Laplace, Uniform and Cauchy perturbations at the same intensity of $0.1$. Gaussian noise achieves the lowest or near-lowest errors in most scenarios, whereas the other distributions show less consistent performance across attacks. Therefore, Gaussian noise with a standard deviation of $0.1$ is adopted as the default perturbation setting, providing the best overall balance between robustness enhancement and nominal behavior preservation.

\section{Limitations and Future research}\label{future research}

Despite the promising results, several limitations remain. First, the evaluation is primarily based on IDM simulations calibrated using ICE-ACC data, and the effectiveness of TRACER in real-world ACC and V2X environments remains to be validated. Second, the current experiments focus on pre-defined cyberattacks involving a single ACC vehicle. Future work will examine coordinated multi-vehicle and previously unseen attack scenarios to further assess the generalization of the recovery policy. 
Finally, RDCs are currently used for rationality-aware recovery triggering rather than as explicit policy constraints. Incorporating RDC-based regularization into policy optimization may further strengthen the physical consistency and interpretability of recovered trajectories.

\section{Conclusion}\label{conclusion}

This paper proposed TRACER, an RDC-triggered trajectory recovery framework for cyberattack-resilient ACC systems. TRACER combines an interpretable, rationality-aware recovery trigger with cross-branch robust imitation learning to address attack-induced distribution shifts. RAT activates recovery when the observed driving state violates the calibrated RDC bounds, whereas CRIL combines clean-branch and noise-enhanced imitation learning with cross-branch refinement to
recover compromised trajectories toward the attack-free driving-state distribution. Experiments across six cyberattack scenarios demonstrate reliable recovery triggering and strong overall performance in recovery accuracy, traffic stability, and safety. State-space analysis further shows that TRACER mitigates attack-induced distribution shifts and restores the coupled spacing–velocity dynamics rather than merely reconstructing individual trajectory states.

\section*{Declaration of generative AI and AI-assisted technologies in the writing process}

During the preparation of this work, the authors used ChatGPT (OpenAI) in order to improve the grammar and writing. After using this tool, the authors reviewed and edited the content as needed and take full responsibility for the content of the manuscript.

\clearpage
\newpage
\bibliographystyle{IEEEtran}

\bibliography{refs}
\end{document}